\documentclass[twocolumn,aps,preprintnumbers,pra,showpacs,amsfonts,amsmath,amssymb,superscriptaddress,floatfix,letterpaper]{revtex4}

\usepackage[dvips]{graphicx}
\usepackage[tight]{subfigure}
\usepackage{times}

\allowdisplaybreaks[3]

\newcommand{\bra}[1]{\langle{#1}|}
\newcommand{\ket}[1]{ |{#1} \rangle}
\newcommand{\bracket}[2]{\langle{#1}|{#2}\rangle}
\newcommand{\bradmket}[3]{\langle #1 | #2 | #3 \rangle}
\newcommand{\expectation}[1]{\langle #1 \rangle}

\newcommand{\RS}{\mathrm{S}}
\newcommand{\RP}{\mathrm{P}}
\newcommand{\ox}{\hat{x}}
\newcommand{\op}{\hat{p}}
\newcommand{\stp}[1]{\mathrm{(#1)}}
\newcommand{\xerr}{x_{\mathrm{error}}}
\newcommand{\pba}{p_{\mathrm{back\ action}}}

\newcommand{\affA}{%
    Department of Applied Physics, School of Engineering,
        The University of Tokyo,\\
    7-3-1 Hongo, Bunkyo-ku, Tokyo 113-8656, Japan}

\newcommand{\affB}{%
Department of Optics, Palack\'{y} University, 17. listopadu 1192/12, 772 07 Olomouc, Czech Republic
}
\begin{document}

\title{Reversing Quantum Nondemolition Interaction as Quantum Erasing}

\date{\today}

\author{Yoshichika Miwa}
\affiliation{\affA}
\author{Jun-ichi Yoshikawa}
\affiliation{\affA}
\author{Ryuji Ukai}
\affiliation{\affA}
\author{Radim Filip}
\affiliation{\affB}
\author{Akira Furusawa}
\affiliation{\affA}


\begin{abstract}
We reverse a quantum nondemolition (QND) interaction and restore the signal quantum state by measurement and feedforward.
This operation corresponds to quantum erasing for continuous variables (CVs).
CV quantum eraser restores the coherence of the signal quantum state by erasing the signal information leaking to another system,
where the information leaking is induced by the QND interaction.
We employ a homodyne measurement for erasing of the information.
Then, by performing a feedforward displacement operation, we restore the initial quantum state together with its coherence.
For verification, we use a coherent state and a squeezed vacuum state as inputs, and then restore one of them or the other, whichever we choose. Experimental results are shown as Wigner functions, average fidelity, change in coherence, and trade-off between leakage information and coherence.
\end{abstract}


\maketitle


\section{Introduction}

Quantum erasing is a historical issue in discussion about quantum complementarity and reversibility of decoherence~\cite{Scully.pra}.
In traditional double-slit-based experiment, the complementarity is written as a trade-off between which-way information and fringe visibility~\cite{trade_off1,trade_off2}.
Here which-way information is non-demolitionably transferred to another particle i.e. measurement probe, while this information collapses the superposition of propagating two different ways, and thus, the fringe disappears.
Generally speaking, arbitrarily interaction between a quantum system and its environment induces decoherence.
Nonetheless, in the case of non-demolishing interaction, proper measurement of the environment can, in principle, completely erase the information and perfectly reverse the decoherence.
Until now, most of quantum erasers are proposed and demonstrated with qubits~\cite{Scully.pra, Kim99.prl}.
In the case of qubits, controlled-NOT operation is utilized as the non-demolition interaction.
However, in most of qubit eraser experiments, controlled-NOT operation is simulated by utilizing entangled photons as input quantum system and its environment.
So, decoherence is not actually induced by the interaction.
Very recently, the quantum eraser which separates input state preparation and interaction was achieved~\cite{Barbieri.njp}.
On the other hand, continuous-variable (CV) quantum erasing reverses the decoherence induced by a quantum non-demolition (QND) interaction~\cite{Radim02.ipp, Filip03.pra}.
The advantage of CV is that the process works deterministically.
An interesting property of these quantum erasers is that coherence or state restoration does not depend on a state of the environment (even arbitrary noisy)~\cite{Schwindt}.
No special condition nor any pre-processing of the environment and the system are required to approach the perfect reconstruction.

A typical discrete-variable qubit eraser works as follows.
Here ``signal'' and ``probe'' are a quantum system and its environment, which are denoted by the subscripts ``S'' and ``P'', respectively.
Suppose the initial signal state is $\ket{\pm}_\RS$, i.e.\ either $\ket{+}_\RS$ or $\ket{-}_\RS$, where $\ket{\pm} = \frac{1}{\sqrt{2}}(\ket{0} \pm \ket{1})$ is a superposition of $\ket{0}$ and $\ket{1}$ which are orthogonal computational-basis (CB) eigen states.
Then, the signal qubit is entangled with the probe qubit in $\ket{0}_\RP$ or alternatively $\ket{1}_\RP$ via a controlled-NOT operation \cite{cnot}.
The resulting state is a fully entangled state $\ket{\Psi}_{\RS \RP} =\frac{1}{\sqrt{2}} (\ket{0}_\RS \ket{0}_\RP \pm \ket{1}_\RS \ket{1}_\RP)$, or alternatively $\ket{\Psi}_{\RS \RP} =\frac{1}{\sqrt{2}} (\ket{0}_\RS \ket{1}_\RP \pm \ket{1}_\RS \ket{0}_\RP)$.
Now the state of signal qubit is decohered and becomes a fully mixed state when we consider only the signal qubit.
Here the density operator is $\hat{\rho}_\RS = \frac{1}{2}(\ket{0}_\RS \bra{0} + \ket{1}_\RS \bra{1})$ for all the cases above, which is derived via tracing out the probe qubit.
The controlled-NOT operation can be interpreted as perfectly copying the signal which-CB ``information'' to the probe qubit.
This {\em leaking information} collapses the superposition, even without any measurement, and the resulting signal qubit state is incoherent mixture of $\ket{0}_\RS$ and $\ket{1}_\RS$.
By erasing the leaking information, we can reverse the decoherence.
This is done by making a measurement in a superposition basis $\ket{\pm}_\RP$ because the measurement reveals the {\em complementary information} and disenables us to access the leaking information.
The measurement results in preserving superposition in the signal qubit as $\ket{\pm}_\RS$ in the case of $\ket{+}_\RP$, or $\ket{\mp}_\RS$ in the case of $\ket{-}_\RP$.
In order to restore the initial qubit states $\ket{\pm}_\RS$, we perform a feed-forward correction, namely making unitary transformation $\ket{\pm}_\RS \to \ket{\mp}_\RS$ if the state $\ket{-}_\RP$ has appeared.
Note that this feedforward does not depend on whether the initial probe state is $\ket{0}_\RP$ or $\ket{1}_\RP$.
Generally speaking, the erasing procedure is universal, i.e., it works for any unknown input states of signal and probe, even if there is no entanglement after the controlled-NOT operation~\cite{Schwindt}.

Mechanism of a CV eraser well corresponds to that of the qubit eraser.
Instead of $\ket{0}$ and $\ket{1}$ for a qubit, coordinate eigenstates $\ket{x}$ for any real number $x$ are the CB in the CV case.
Suppose initial signal state is $\ket{\psi}_\RS = \int dx \psi(x) \ket{x}_\RS$ as a superposion of the CB.
Then, the signal mode is entangled with the probe mode by a QND interaction, where its unitary operator is $\hat{U}=\exp (i\ox_\RS \op_\RP/\hbar)$.
Similary as for the qubit case, the QND interaction transfers information only in the single variable.
Although initial probe state is arbitrary for CV quantum erasing, we assume $\ket{x=0}_\RP$ (denoted as $\ket{0}_\RP$ for simplicity) here.
Later, we describe the general case.
In the case of $\ket{0}_\RP$, the output state is $\ket{\Psi}_{\RS \RP} = \int dx \psi(x)\ket{x}_\RS \ket{x}_\RP$.
The signal-coordinate {\em information} is completely copied to the probe by the QND interaction.
Thus, the state of the signal mode becomes a mixed state whose density operator is $\hat{\rho}_\RS = \int dx |\psi(x)|^2 \ket{x}_\RS \bra{x}$.
Information transfer to the probe mode in the coordinate basis collapses the superposition.
In the case that we access the {\em information} by making a measurement on the coordinate basis of the probe, the signal state becomes a coordinate eigen state $\ket{x}_\RS$, where the measurement is known as a QND measurement.
In contrast, measurement on the momentum basis, namely $p$, erases the {\em leaking information} and the coherence of the signal mode is restored.
This is because the coordinate and momentum operators are conjugate to each other.
The eraser can also restore the unknown initial state $\ket{\psi}_\RS$ with proper feedforward,
while experimental qubit erasers are usually evaluated only on recovery of the coherence~\cite{Scully.pra,Kim99.prl,Barbieri.njp}.

This CV quantum erasing can be regarded as undoing QND interaction with measurement and feedforward.
The unitary operator of the QND interaction is $\exp (i\ox_\RS \op_\RP/\hbar)$, so the interaction is reversible by another QND interaction i.e. $\exp (-i\ox_\RS \op_\RP/\hbar)$,
which can be decomposed into measurement $\op_\RP \to p_0$ and feedforwarding phase space displacement $\exp (-i\ox_\RS p_0/\hbar)$.
Thus, quantum erasing corresponds to inverse operation of the QND interaction based on the measurement and assisted by classical communication.
From this reason, quantum erasing works equally well with any input states, in principle.

In the case of experiment, however, QND interaction always contains some imperfection.
In our implementation of QND interaction~\cite{Filip05.pra,Yoshikawa08.prl}, the gain $g$ of the QND interaction $\exp (ig\ox_\RS \op_\RP/\hbar)$ can be set to unity.
On the other hand, there is some excess noise which originates from ancillary squeezed states utilized as resource for the nonclassicality of the interaction.
This excess noise corresponds to imperfection in concealing the information which leaks to ancilla.
Through the erasing process, the leaking information in the probe is erased, but that in the ancilla can not be removed.
So the excess noise contaminating the restored state originates from the QND interaction and does not depend on either the initial signal or probe state.
In this sense, both QND interaction and quantum erasing works equally well with any set of input states.

The first experimental CV quantum eraser was reported in Ref.~\cite{Andersen04.prl}.
In the experiment, the leaking information in the probe is partly erased.
However, it was not demonstration of reversing QND interaction, a beam splitter interaction is utilized instead.
Since beam splitter interaction is not a non-demolition interaction, both coordinate and momentum of the probe hold the leaking information.
From this reason, the probe state is limited to a squeezed state, whose anti-squeezed noise conceals the leaking information of momentum variable and thus permits erasing.
In their scheme, the probe also works as resource for the nonclassicality, not the whole leaking information in the probe is erased without infinite squeezing, and thus restored state depends on the probe.
Without probe squeezing, the quantum erasing with beam splitter interaction requires pre-processing of the unknown input state~\cite{Filip09} or measurement of both of the variables, which leads only to limited reconstruction of the input state~\cite{Sabuncu}.
In their experiment, performance for other than vacuum state is not demonstrated.
Moreover, the change in the coherence is not directly observed.
We reported cluster state shaping as application of quantum erasing in Ref.~\cite{shaping}.
This technique provide flexibility of fixed large-scale cluster states and enable us to shape it into modified, smaller cluster states~\cite{shaping,one-wayGu}.
Nonetheless, in that experiment, we also utilized beam splitters instead of QND interactions for cluster state generation, and we dealt only with the cluster state and its resource squeezed vacuum states.

In this paper, we demonstrate CV quantum erasing as reversing a QND interaction.
With a QND interaction, we can restore any initial signal state, and the resulting state does not depend on the initial probe state.
First, we use a coherent and squeezed-vacuum states as the initial signal and probe states, respectively.
We verify the change in the quantum states and coherence throughout the process by performing homodyne tomography.
Then, in order to verify the imperfection independence from the initial states, we exchange the role of input states, i.e.\ we utilize a squeezed vacuum state as initial signal state and the coherent state as the probe.
Finally, we also verify information erasure by uncertainty relation between information and decoherence.

\renewcommand*{\thesubfigure}{(\roman{subfigure})}
\begin{figure}[tb]
\centering
\subfigure[CV quantum erasing.]{
\includegraphics[clip, scale=0.4]{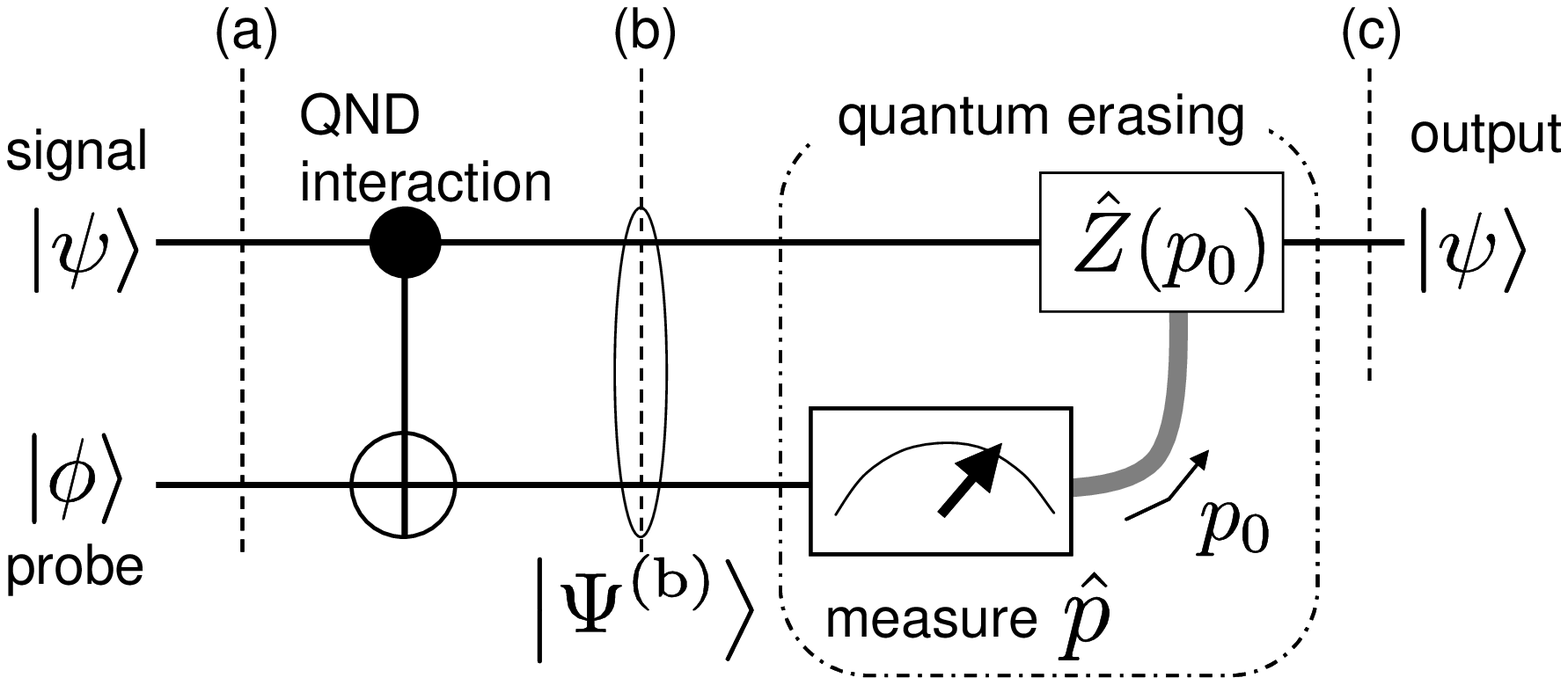}
\label{fig:erasing}
}
\subfigure[QND measurement.]{
\includegraphics[clip, scale=0.4]{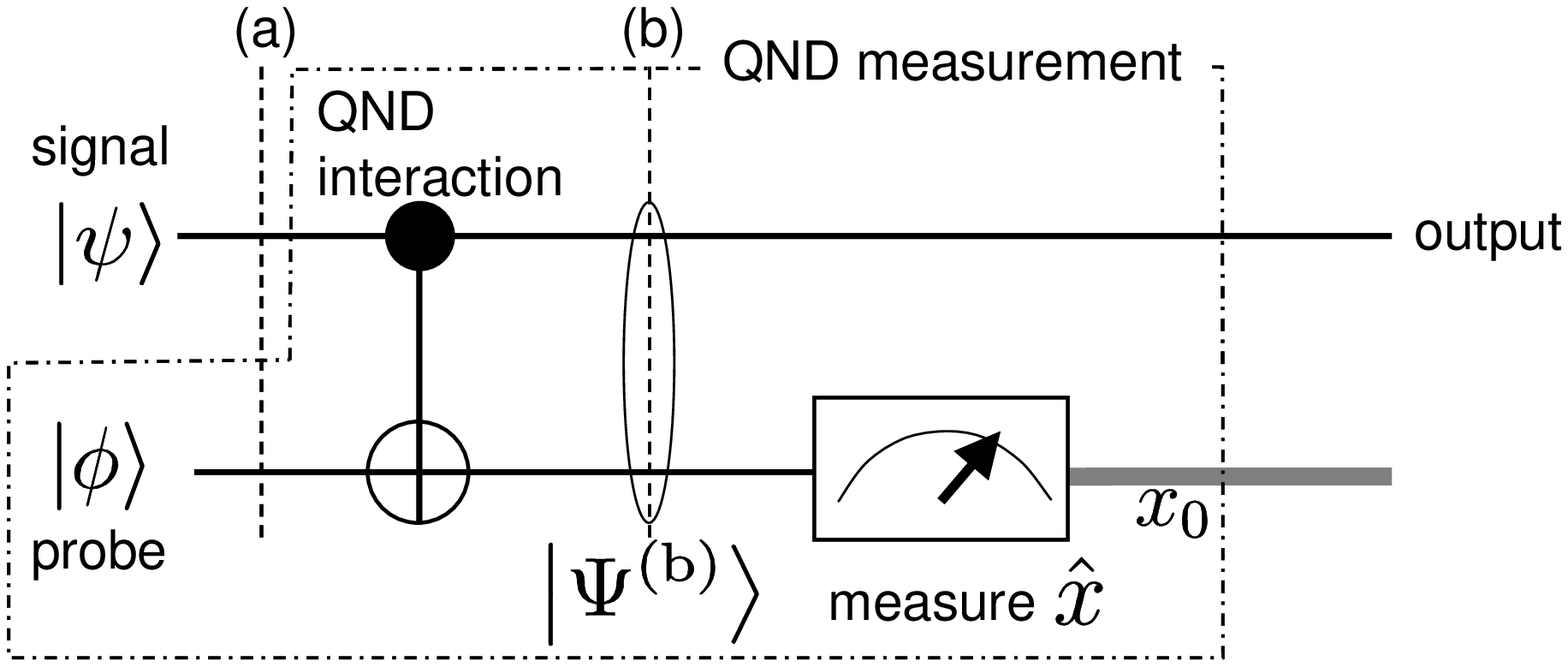}
\label{fig:qndm}
}
\subfigure[Experimental setup for the CV quantum erasing.]{
\includegraphics[clip, scale=0.3]{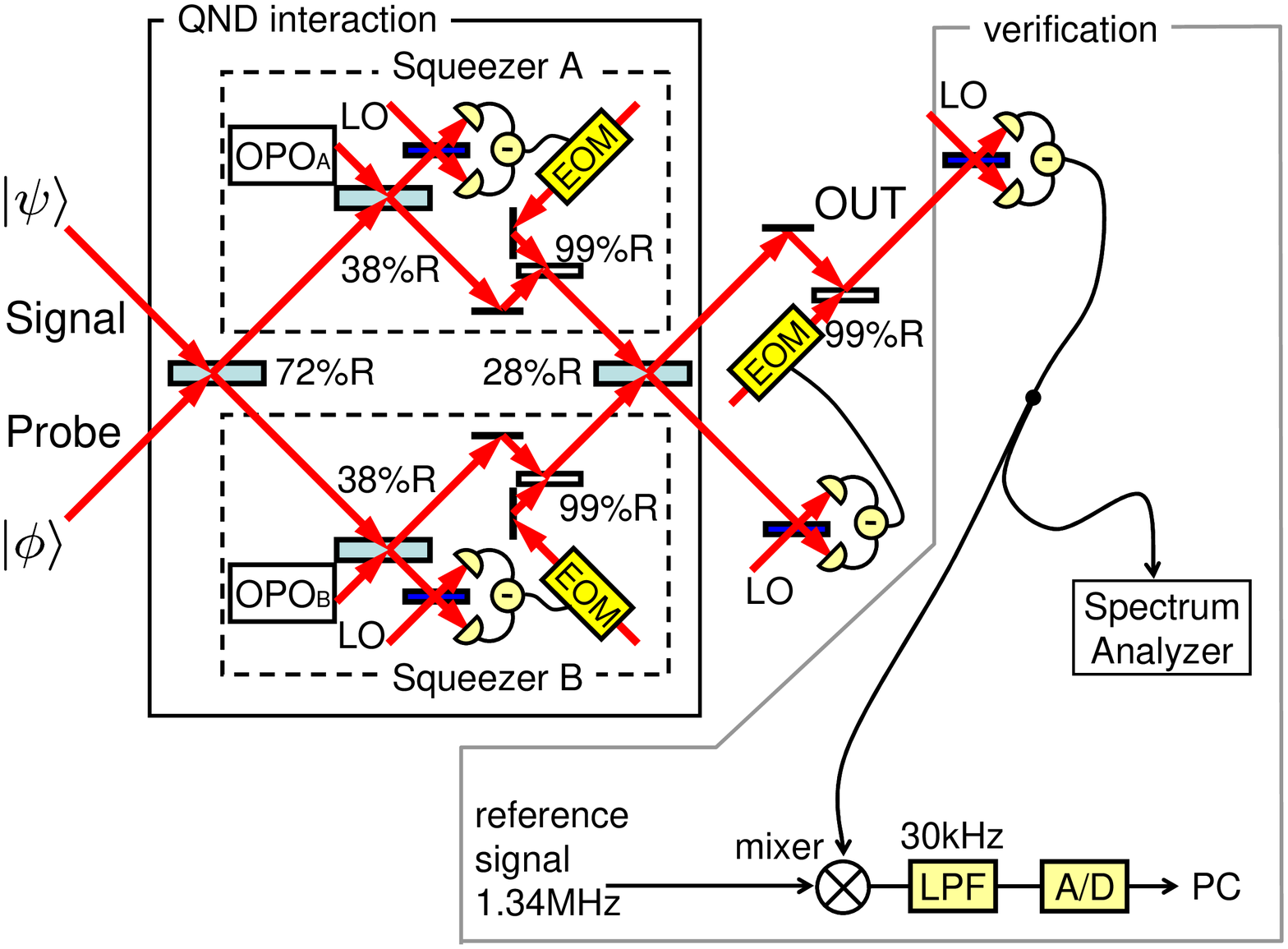}
\label{fig:ex_erasing}
}
\caption{(Color online) Schematic and our optical setup of CV quantum erasing.
OPO: optical parametric oscillator, LO: optical local oscillator, and EOM: electro-optic modulator.}
\end{figure}


\section{Theory}
In this section, we describe the quantum states throughout the process and discuss the relation between the {\em leaking information} and decoherence which are induced by QND interaction.
Figure~\ref{fig:erasing} shows a schematic of CV quantum erasing.
Input quantum states of signal and probe modes are independently prepared (a), and then a QND interaction couples them (b), finally, measurement and feedforward restore the initial signal state (c).
Hereafter, quantum states in these steps are denoted by superscripts or labels (a), (b), and (c) and they are normalized with $\hbar = 1/2$.

$\ket{\psi}_\RS$ and $\ket{\phi}_\RP$ represent arbitrary initial quantum states of the signal and probe modes, respectively.
For ease of explanation, we assume pure states as initial quantum states.
Note that we discuss the case of mixed state later.
By QND interaction, these modes are entangled as follows,
\begin{align}
\ket{\Psi^\stp{b}} &= e^{2i\ox_\RS \op_\RP}\ket{\psi}_\RS \ket{\phi}_\RP \notag \\
&= \iint dx_\RS dx_\RP \ket{x_\RS}_\RS \ket{x_\RP +x_\RS}_\RP \psi (x_\RS)\phi (x_\RP) \label{eq:QNDx} \\
&= \iint dp_\RS dp_\RP \ket{p_\RS -p_\RP}_\RS \ket{p_\RP}_\RP \tilde{\psi} (p_\RS)\tilde{\phi} (p_\RP), \label{eq:QNDp}
\end{align}
where $\psi (x_\RS) \equiv \bracket{x_\RS}{\psi}$ and $\phi (x_\RP) \equiv \bracket{x_\RP}{\phi}$ are the input wave functions on coordinate bases of the signal and probe, respectively,
while $\tilde{\psi} (p_\RS) \equiv \bracket{p_\RS}{\psi}$ and $\tilde{\phi} (p_\RP) \equiv \bracket{p_\RP}{\phi}$ are the input wave functions on momentum bases,
which are obtained by the Fourier transformation denoted by ``\~\ '' via $\bracket{x}{p} = e^{i2xp}/\sqrt{\pi}$.
Obviously, this entangled state has correlation in both coordinate and momentum variables.

The following density matrices represent the each mode state after the QND interaction derived from Eq.~\eqref{eq:QNDx} and Eq.~\eqref{eq:QNDp} by tracing out the other mode:
\begin{align}
\rho_\RS^\stp{b}(x,x^\prime ) &= \rho_\RS (x,x^\prime) \int d\xi \rho_\RP (x^\prime -x+\xi ,\xi ), \label{eq:QND0} \\
\tilde{\rho}_\RS^\stp{b}(p,p^\prime ) &= \int d\eta \tilde{\rho}_\RS (p+\eta ,p^\prime +\eta ) \tilde{\rho}_\RP (\eta ,\eta ), \label{eq:QND1} \\
\rho_\RP^\stp{b}(x,x^\prime ) &= \int d\xi \rho_\RP (x-\xi ,x^\prime -\xi )\rho_\RS (\xi ,\xi), \label{eq:QND2} \\
\tilde{\rho}_\RP^\stp{b}(p,p^\prime ) &= \tilde{\rho}_\RP (p,p^\prime ) \int d\eta \tilde{\rho}_\RS (p-p^\prime +\eta ,\eta ), \label{eq:QND3}
\end{align}
where $\rho_\RS, \rho_\RP; \tilde{\rho}_\RS, \tilde{\rho}_\RP$ represent initial density matrices of signal and probe on coordinate bases, and on momentum bases, respectively, e.g.\ $\rho_\RS(x,x^\prime )=\psi (x) \psi^\ast (x^\prime )$, and $\rho_\RS, \dots$ with superscript $\stp{b}$ represent density matrices after the QND interaction.
Equations (\ref{eq:QND0}-\ref{eq:QND3}) are also satisfied with mixed state inputs, and thus we deal with arbitrary input states hereafter.

Among these equations, Eq.~\eqref{eq:QND2} represents that the probe state has the information of the signal coordinate.
We can obtain the information by making a measurement of the probe coordinate.
The measurement outcome $x_0$ is obtained with the probability being a convolution $\rho_\RS(x_0,x_0) \circ \rho_\RP(x_0,x_0)$, where the probability is calculated by applying $x=x^\prime =x_0$ to Eq.~\eqref{eq:QND2}.
This measurement is known as the QND measurement [Fig.~\ref{fig:qndm}].
The measurement outcome $x_0$ contains the error which is the convoluted distribution of $\rho_\RP(x,x)$ originating from the probe initial state.
So the variance of the measurement outcome is the sum of those of the initial signal coordinate and measurement error i.e.\ $( \Delta x^\stp{b}_\RP )^2 = ( \Delta x^\stp{a}_\RS )^2 + (\Delta \xerr )^2$.
The ambiguity of information can be defined as the standard deviation of the measurement error $\Delta \xerr $ which equals to that of the initial probe coordinate $\Delta x^\stp{a}_\RP$.

As quantitative measure of the information, we use Fisher information~\cite{Fisher}, denoted as $I$.
If the Fisher information increases $N$-times, only $1/N$ times of repeated measurements are enough to achieve the same quality of estimation.
It allows us on a physical basis to quantitatively compare any possible change of {\em information} accessible by the QND measurement.
In the case of Gaussian added noise in measurement without a priori knowledge, it satisfies $I=\frac{1}{(\Delta x_{\mathrm{error}})^2}$ (see appendix).

The Fisher information decoheres the signal state as represented in Eq.~\eqref{eq:QND0}.
The function $R(x^\prime -x)=\int d\xi \rho_\RP (x^\prime-x+\xi ,\xi )$ is the ratio of each density matrix elements, i.e.\ $R(x^\prime -x)=\frac{\rho_\RS^\stp{b}(x,x^\prime )}{\rho_\RS (x,x^\prime)}$.
The ratio function is equal to $\int dp e^{i2(x^\prime -x)p}\tilde{\rho}_\RP(p,p)$, and thus it equals to $1$ in the case of $x=x^\prime $ because of the normalization of $\tilde{\rho}_\RP$.
On the other hand, it satisfies $|R(x^\prime-x)| \leq 1$ in the case of $x\not= x^\prime$.
So the diagonal elements of $\rho_\RS$ (signal density matrix on the coordinate basis) are preserved through the QND interaction, while the off-diagonal elements ($x\not= x^\prime$) become smaller in absolute value with the ratio of $R(x^\prime -x)$.
Thus, the distribution of $x_\RS$ is not demolished, while the superposition between any different coordinate basis states $\ket{x}_\RS$ and $\ket{x^\prime}_\RS$ is decohered.
It well corresponds to double-slit-based qubit cases, where the fringe reflects the single off-diagonal element $\rho(0,1)$.
The which-way (0 or 1) information makes the off-diagonal element smaller.
Here, not only a single off-diagonal element, but rather a continuous set of the off-diagonal elements describes the coherence in CV system.

In momentum basis, this decoherence appears as the increase of variance as shown in Eq.~\eqref{eq:QND1}, where the distribution of momentum $p$ is calculated to be $\tilde{\rho}_\RS(p,p) \circ \tilde{\rho}_\RP(-p,-p)$.
In the theory of quantum measurement~\cite{Braginsky}, this increase of variance corresponds to back action of the measurement, i.e.\ the back action increases as measurement error decreases.
The trade off satisfies the following uncertainty relation of measurement~\cite{Braginsky},
\begin{align}
\Delta \xerr \Delta \pba &\geq \frac{1}{4},
\label{eq:uncertainty of measurement}
\end{align}
where $\Delta \pba $ denotes the standard deviation of the back action which is the convoluted additional distribution in the signal momentum after the interaction,
i.e.\ $(\Delta \pba )^2 = (\Delta p^\stp{b}_\RS )^2 - (\Delta p^\stp{a}_\RS )^2 $.
In this case, the backaction originates from the initial probe, i.e.\ $\Delta \pba = \Delta p^\stp{a}_\RP$.
Thus, we have equality in Eq.~\eqref{eq:uncertainty of measurement} if and only if the probe is in minimal uncertainty state $\Delta x^\stp{a}_\RP \Delta p^\stp{a}_\RP = 1/4$.
From the inequality, the minimal back action derived as $\Delta \pba^{\mathrm{min}} = 1/(4\Delta \xerr )$.

As quantum erasing operation, we employ the measurement of $\op_\RP$, hereafter $p_0$ represents the measurement outcome.
By applying $p=p^\prime =p_0$ to Eq.~\eqref{eq:QND3}, we calculate the distribution of the measurement outcome is $|\tilde{\rho}_\RP (p_0,p_0)|^2$.
Thus, the measurement result does not reflect any property of the initial signal state.
Besides, the measurement of $\op_\RP$ erases the leaking information from $\ox_\RP$ owing to their conjugateness.
Therefore, the decoherence of the signal disappears together with the leaking information.
The signal density matrix after the measurement is $\rho_\RS (p,p)|_{p_0} = \rho_\RS (p-p_0,p-p_0)$.
Thus, the initial signal state is restored by a feedforwarding momentum-displacement operation by $p_0$, i.e.\ $\hat{Z}_\RS (p_0)= \exp (-i2\ox_\RS p_0)$.

With a perfect quantum erasing operation, the back action is completely removed, i.e.\ $\Delta p_\RS^\stp{c} = \Delta p_\RS^\stp{a}$, which corresponds to vanishing the Fisher information.
In experiments, there is some residual noise $(\Delta p_{\mathrm{residual\ noise}})^2 = (\Delta p_\RS^\stp{c})^2 - (\Delta p_\RS^\stp{a})^2$ after the erasing operation due to imperfections of experimental QND interaction~\cite{Yoshikawa08.prl}. 
Nonetheless, the leaking information should be partly erased if the residual noise is below the minimal back action, i.e., $\Delta p_{\mathrm{residual\ noise}} < \Delta p_{\mathrm{back\ action}}^{\mathrm{min}}$, or equivalently,
\begin{align}
\Delta \xerr \Delta p_{\mathrm{residual\ noise}} &< \frac{1}{4}.
\label{eq:uncertainty of measurement2}
\end{align}
We consider inequality \eqref{eq:uncertainty of measurement2} as a sufficient condition for the success of information erasure. Quantitatively, the maximal residual Fisher information $I_{\mathrm{residual}}=16\Delta^2 p_{\mathrm{residual\ noise}}$ after erasing can be compared with the Fisher information $I$ after the QND measurement.
Note that this relation does not conflict with inequality \eqref{eq:uncertainty of measurement} because the erasure of leaking information corresponds to making the erorr in the QND measurement infinity. $\Delta \xerr $ represents measurement error without erasing, and thus $\Delta \xerr $ and $\Delta p_{\mathrm{residual\ noise}}$ cannot be obtained simultaneously.

Then we consider the case with experimental imperfect QND operation~\cite{Yoshikawa08.prl}.
The transfer gains (coefficient of amplitude transfer) of the QND gate can slightly vary owing to change in optical interference visibilities.
There are excess Gaussian noises on both $x$ and $p$ quadratures coming from ancillary modes utilized as resource for entanglement.
By taking acount of these imperfections, density operator after the erasing is rewritten as follows:
\begin{align}
\rho_{\RS \frac{\pi}{4}}^\stp{c}\left( \frac{x+x^\prime}{\sqrt{2}} ,\frac{x-x^\prime}{\sqrt{2}} \right) &\propto
\int d\xi \rho_{\RS \frac{\pi}{4}} \left( \frac{x+x^\prime -\xi}{\sqrt{2}g_x}, g_p\frac{x-x^\prime}{\sqrt{2}}\right) \notag \\
& \! \! \! \! \! \! \! \! \! \! \! \! \! \! \times \exp \left[-\frac{\xi^2}{2\sigma_x^2}\right] \exp \left[-\frac{(x-x^\prime )^2}{2\sigma_p^2}\right], \\
\text{where\ } \rho_{\frac{\pi}{4}}(\alpha , \beta ) &= \rho \left( \frac{\alpha +\beta}{\sqrt{2}},\frac{\alpha -\beta}{\sqrt{2}}\right) , \notag
\end{align}
and where $g_x$ and $g_p$ denotes the gains in $x$ and $p$ quadratures i.e.\ $\expectation{x}_\RS^\stp{c} = g_x\expectation{x}_\RS^\stp{a}$, $\expectation{p}_\RS^\stp{c} = g_p\expectation{p}_\RS^\stp{a}$; and $\sigma_x^2$ and $\sigma_p^2$ denotes the variance of excess noises in $x$ and $p$ quadratues i.e.\ $(\Delta x_\RS^\stp{c})^2 = g_x^2(\Delta x_\RS^\stp{a})^2 + \sigma_x^2$, $(\Delta p_\RS^\stp{c})^2 = g_p^2(\Delta p_\RS^\stp{a})^2 + \sigma_p^2$, respectively.
These gains and excess noises fully characterize the QND and erasing operation because they are Gaussian operation (i.e. operation whoes Hamiltonian is no more than second order) and its reverse operation, respectively.
The residual noise is the same as the excess noise in $p$ quadrature i.e.\ $\sigma_p^2 = \Delta^2 p_{\mathrm{residual\ noise}}$.
Note that $\xerr$ and $\pba$ also contain excess noises of the QND interaction in experiment.

In our scheme, a restored state does not depend on the probe because the back action from the probe is removed.
This corresponds to full erasure of the leaking information in the probe, even with aforementioned experimental imperfections.
Since our QND and erasing operations are fully Gaussian, the amount of back action from the probe corresponds to the transfer gain from the probe to the signal.
Thus we experimentally estimate the transfer gain by utilizing a coherent state as the probe.


\section{Experimental Setup}
Figure~\ref{fig:ex_erasing} shows our optical implementation of Fig.~\ref{fig:erasing}.
The experimental setup consists of the following parts: preparation of the input signal and probe states (not shown in the figure), QND interaction~\cite{Yoshikawa08.prl}, measurement, feedforward, and finally, the verification measurement.
This setup is similar to that of our quadratic phase gate~\cite{Miwa09.pra}.
However, since the quadratic phase gate is generalized teleportation~\cite{Miwa09.pra}, it requires a squeezed vacuum state as the initial probe state, while the quantum erasing operation does not depend on the initial probe state.
Besides, the mode to be measured and the mode to be suffered from feedforward are also different between these two operations.

As a light source, we utilize a continuous-wave Ti:sapphire laser with the wave length of 860~nm.
As one of two input states, a coherent state at the 1.34~MHz sideband is generated by modulating a weak laser beam of about 10~$\mu$W using two electro-optic modulators (EOMs).
One of the EOMs modulates the phase of the beam and the other modulates the amplitude.
In our optical implementation, coordinate $x$ and momentum $p$ correspond to these amplitude and phase quadratures.
Thus, these two EOMs can generate a coherent state at 1.34~MHz with any complex amplitude from the laser beam.

As the other input state, a squeezed vacuum state is generated by a sub-threshold optical parametric oscillator (OPO).
We also utilize another two OPOs in order to prepare resource squeezed-vacuum states for QND interaction.
Each OPO is a bow-tie shaped cavity of 500~mm in length with a 10-mm-long periodically-poled KTiOPO$_4$ (PPKTP) crystal as a nonlinear medium~\cite{Suzuki06.apl}.
In order to pump the OPOs, we generate second harmonic (430~nm in wavelength) of Ti:sapphire output by a frequency doubler, which has similar structure of OPO while it has KNbO$_3$ crystal instead of PPKTP.
Each generated squeezed state has squeezing level of about $-5$ dB relative to the shot-noise level (SNL).

The QND interaction consists of a Mach-Zehnder interferometer with a single-mode squeezing gate in each arm~\cite{Yoshikawa08.prl}. Each single-mode squeezing gate contains a squeezed vacuum ancilla, homodyne detection, and feedforward~\cite{Filip05.pra, Yoshikawa07.pra}.
The four reflectivities of 72\%, 38\%, 38\%, and 28\% are chosen to achieve unity gain QND interaction as shown in Fig.~\ref{fig:ex_erasing}~\cite{Yoshikawa08.prl}.
These reflectivities are implemented as variable beam splitters (VBSs), each composed of two polarizing beam splitters and a half-wave plate.
This variability enables us to eliminate the QND interaction when we measure input states, by setting the reflectivities of these VBSs to unity.
It is also useful when we exchange the signal/probe roles of two modes, in that case, we flip the two output ports of QND interaction by adjusting the reflectivity of the fourth one from 28\% to 72\% .
Feed-forwarding displacement operation is performed by sending measurement outcome of homodyne detection to an EOM which modulates phase of an auxiliary beam,
and then by coupling the beam to main stream with a 99\% reflector.
At each beam splitter, we lock the relative phase of the two input beams by means of active feedback to a piezoelectric transducer.
For this purpose, two modulation sidebands of 154 and 107~kHz are used as phase references.

To verify the output state, we employ another homodyne detection.
We perform two different verifications.
First, in order to reconstruct the resulting quantum state, we perform optical homodyne tomography,
namely, quantum state reconstruction from the marginal distributions for various phases~\cite{Lvovsky09.rmp}.
We slowly scan the phase of optical local oscillator (LO) and perform a series of homodyne measurements.
The 1.34 MHz component of the homodyne signal is extracted by means of lock-in detection: it is mixed with a reference signal and then sent through a 30~kHz low pass filter,
and then, it is digitized with sampling rate of 300,000 samples per second.
Second, for accurate evaluation of variance in $x$ or $p$-quadrature, we lock the LO phase and extract 1.34 MHz component of the measurement outcome via a spectrum analyzer.
Through this variance evaluation, we verify the information erasure.

The powers of the LOs are about 3~mW.
The detector's quantum efficiencies are greater than 99\%, the interference visibilities to the LOs are on average 98\%, and the circuit noise of each homodyne detector is about 17~dB below the SNL produced by the LO.
Propagation losses of our whole setup are about 7\%.
These losses are compensated via the equality of losses between inputs and outputs.

\renewcommand*{\thesubfigure}{(\alph{subfigure})}
\begin{figure}[tb]
\centering
\subfigure[Input state in each mode.]{
\includegraphics[clip, scale=0.35]{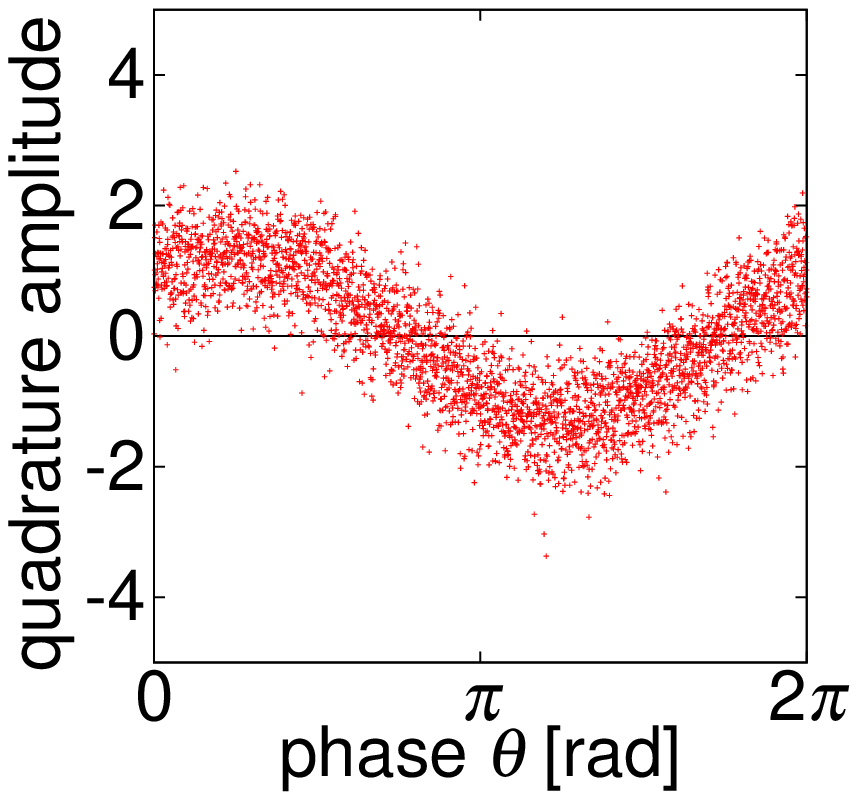}
\includegraphics[clip, scale=0.35]{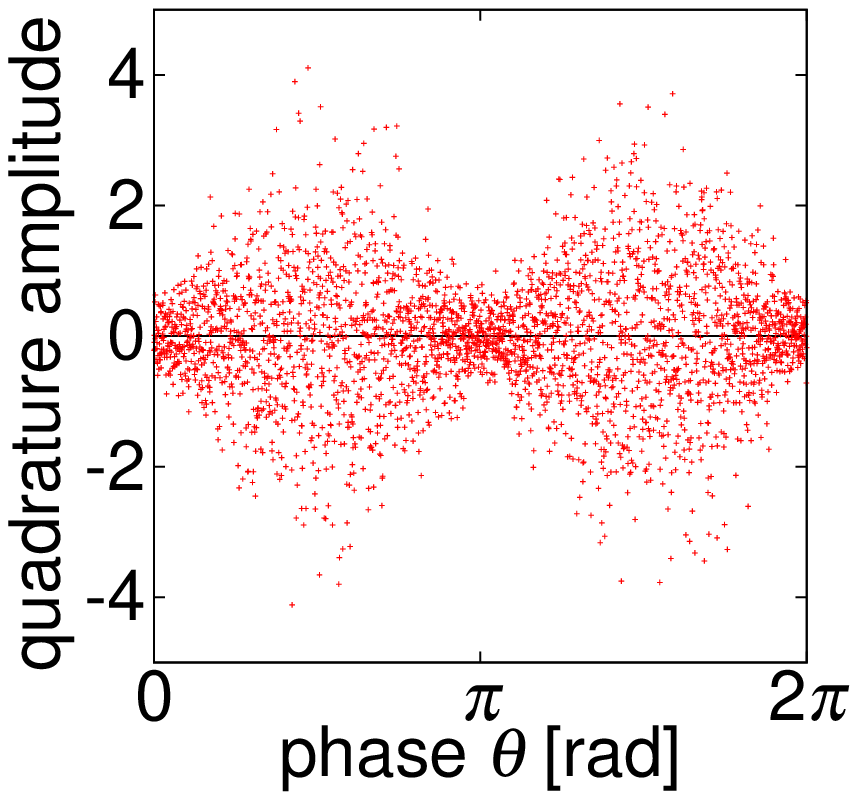}
}
\subfigure[After QND interaction.]{
\includegraphics[clip, scale=0.35]{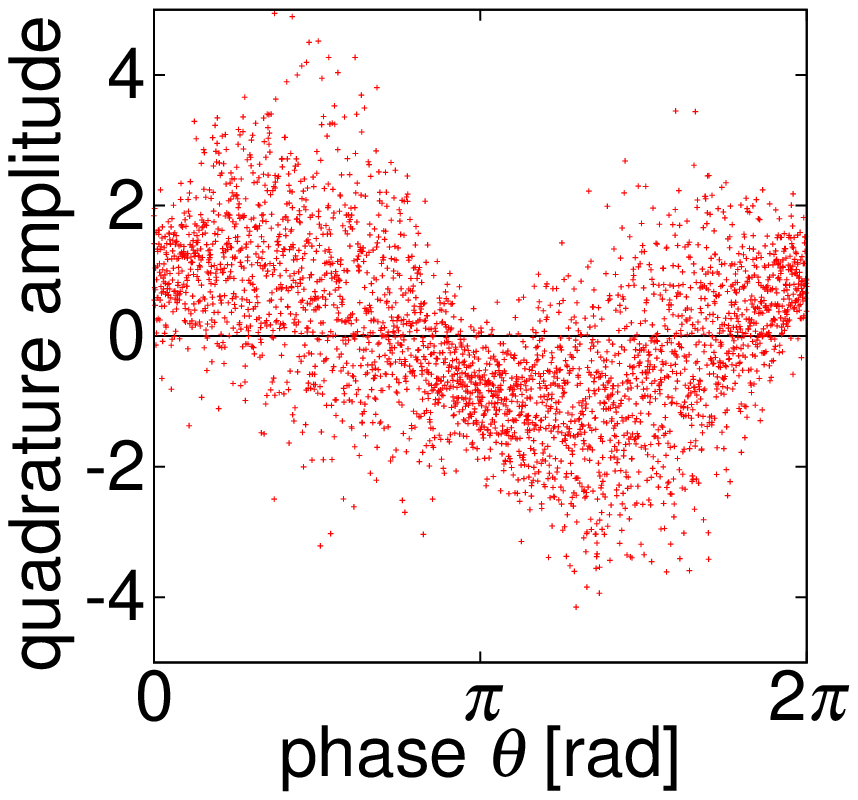}
\includegraphics[clip, scale=0.35]{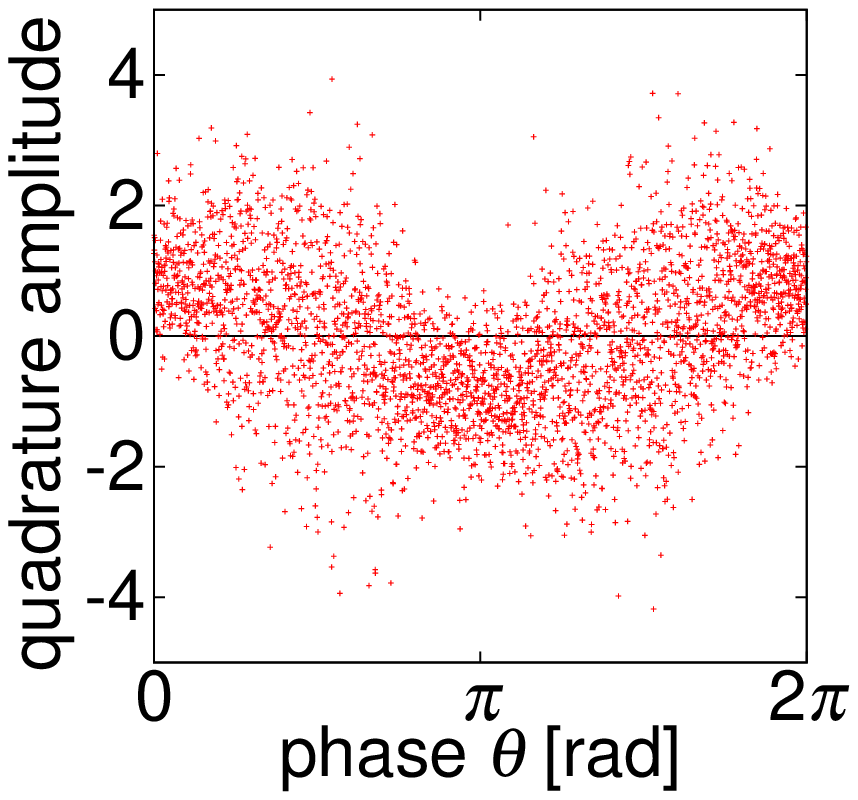}
}
\subfigure[Resulting state of the first mode after~erasing~the~second~mode.]{
\includegraphics[clip, scale=0.35]{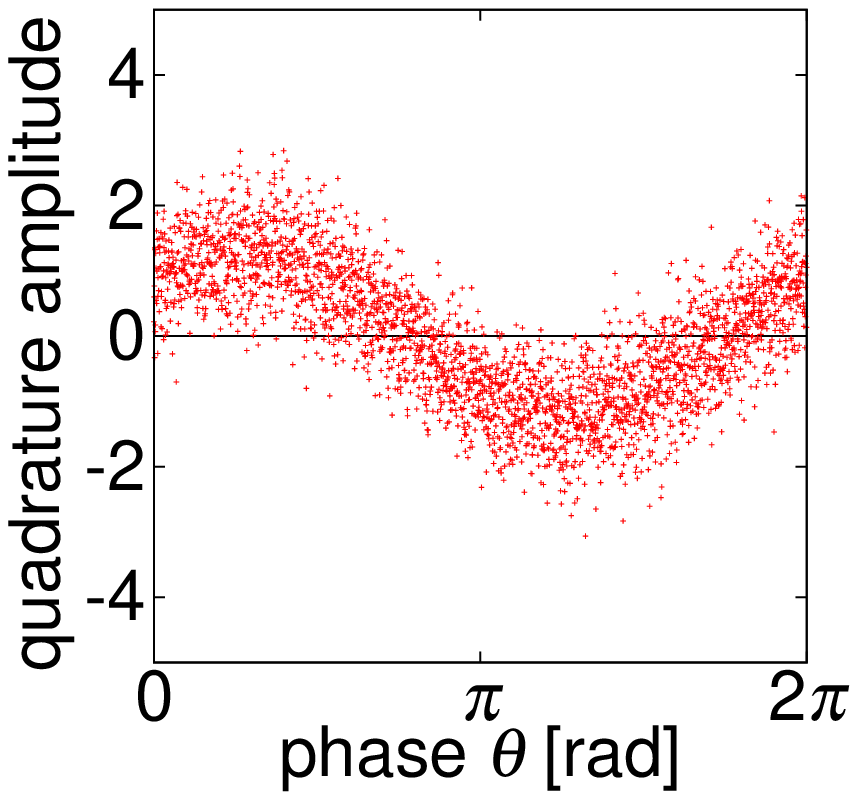}
\includegraphics[clip, scale=0.35]{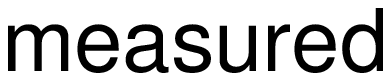}
}
\subfigure[Resulting state of the second mode after~erasing~the~first~mode.]{
\includegraphics[clip, scale=0.35]{measured2}
\includegraphics[clip, scale=0.35]{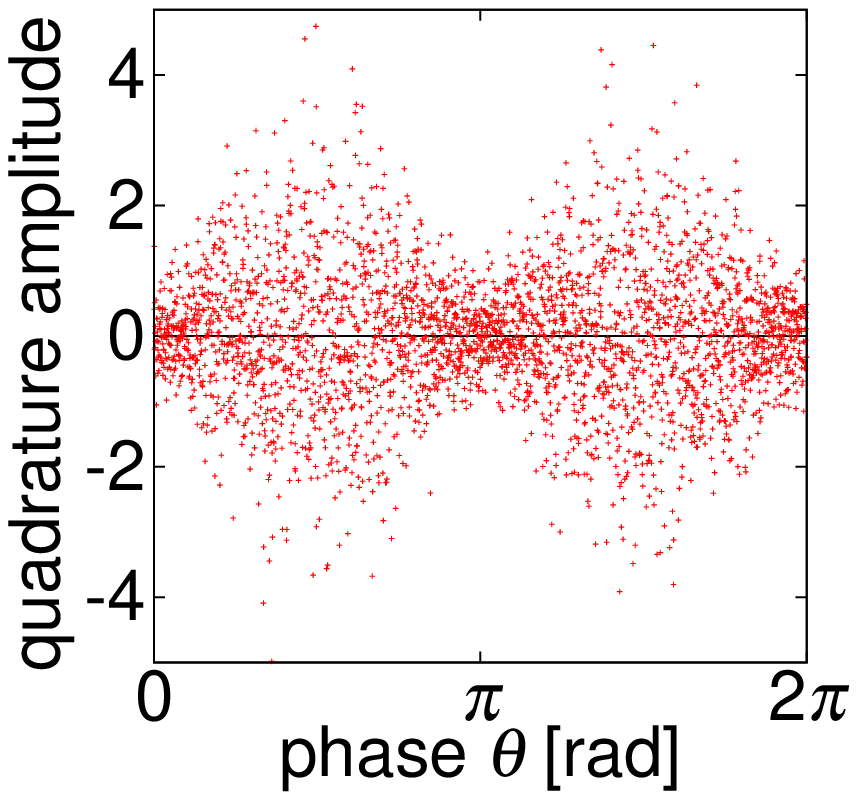}
}
\caption{(Color online) Experimental marginal distribution in each step~/~mode of quantum erasing. Left-side figures are marginal distributions of the first mode and right-side ones are those of the second mode. The input states of the first and second modes are a coherent state and a squeezed-vacuum state, respectively~(a). The QND interaction entangles them, thus each mode is decohered~(b). By erasing the leaking information from second (first) mode, the initial state in the first (second) mode is recovered~(c) [(d)].
Each figure is plotted every 20 points of raw data.}
\label{fig:erasing_2pi}
\end{figure}

\begin{figure}[tb]
\centering
\subfigure[Input state in each mode.]{
\includegraphics[clip, scale=0.4]{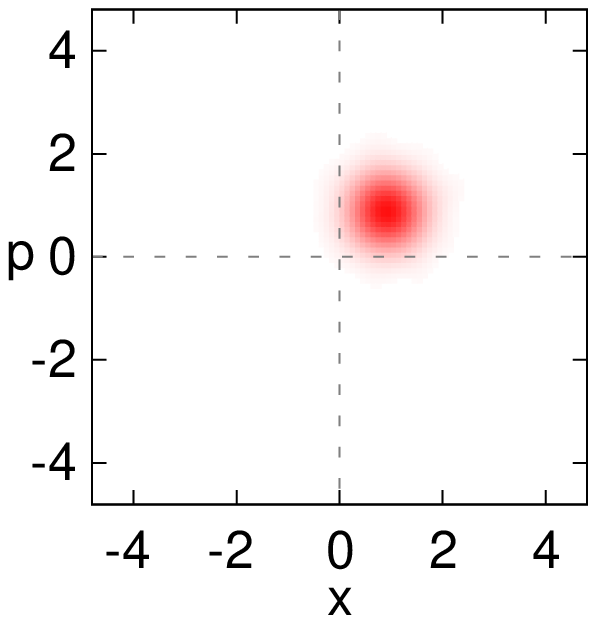}
\includegraphics[clip, scale=0.4]{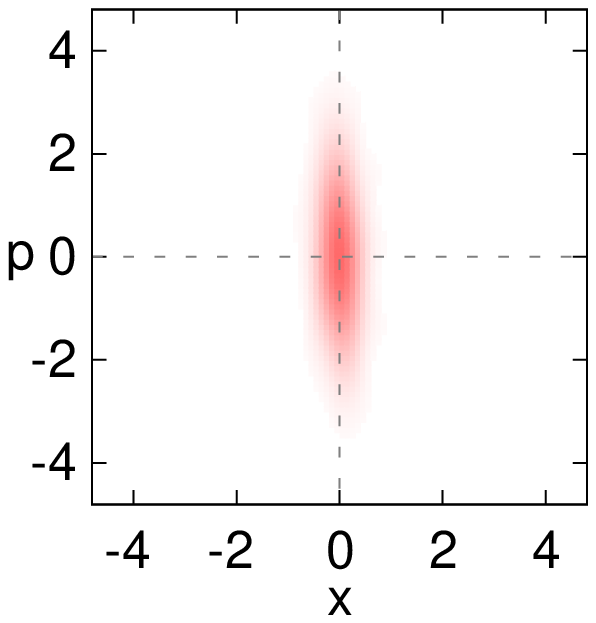}
}
\subfigure[After QND interaction.]{
\includegraphics[clip, scale=0.4]{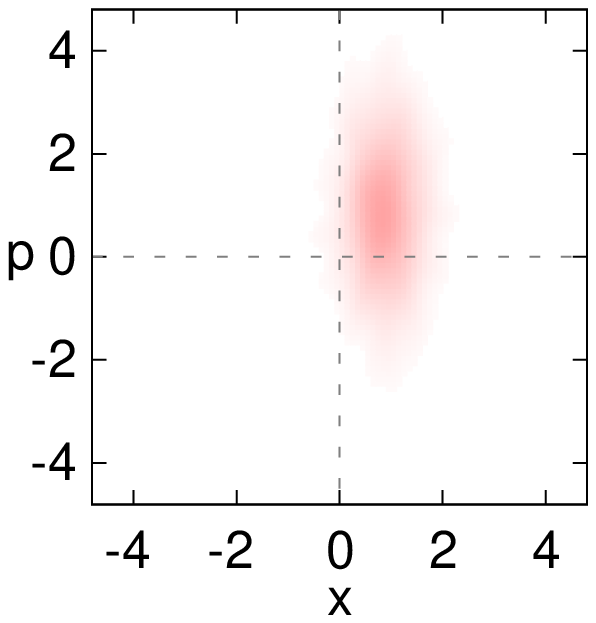}
\includegraphics[clip, scale=0.4]{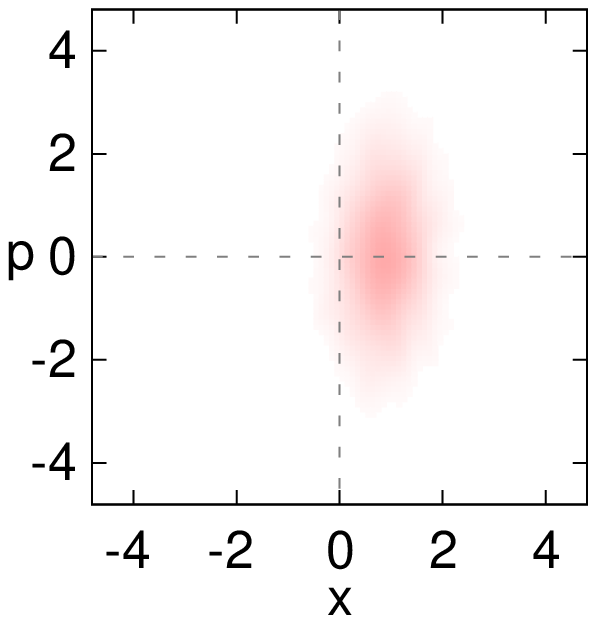}
}
\subfigure[Resulting state of the first mode after~erasing~the~second~mode.]{
\includegraphics[clip, scale=0.4]{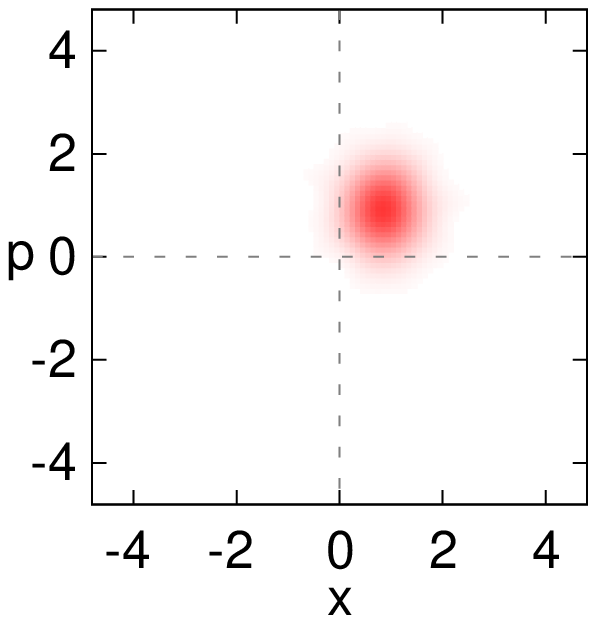}
\includegraphics[clip, scale=0.4]{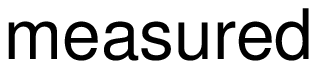}
}
\subfigure[Resulting state of the second mode after~erasing~the~first~mode.]{
\includegraphics[clip, scale=0.4]{measured}
\includegraphics[clip, scale=0.4]{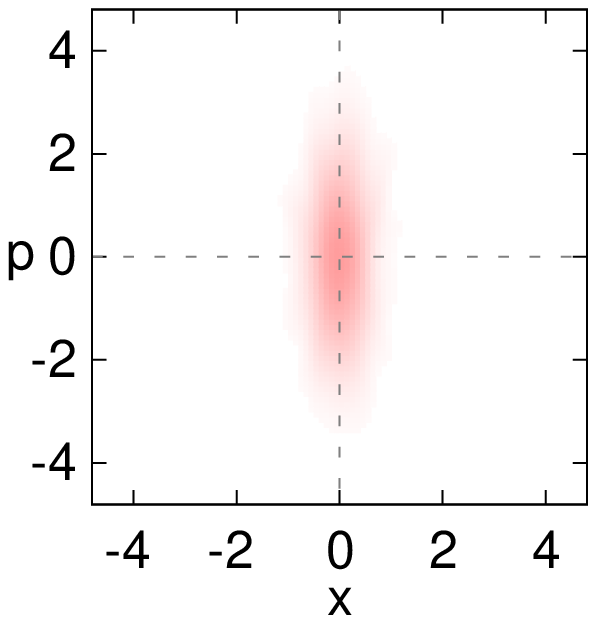}
}
\caption{(Color online) Reconstructed Wigner functions from experimental marginal distributions shown in Fig.~\ref{fig:erasing_2pi}.
Left-side figures are Wigner functions of the first mode and right-side ones are those of the second mode. The input states of the first and second modes are a coherent state and a squeezed-vacuum state, respectively~(a). The QND interaction entangles them, thus each mode is decohered~(b). By erasing the leaking information from second (first) mode, the initial state in the first (second) mode is recovered~(c) [(d)].}
\label{fig:erasing_Wig}
\end{figure}

\begin{figure}[tb]
\centering
\subfigure[Input density matrix.]{
\includegraphics[clip, scale=0.55]{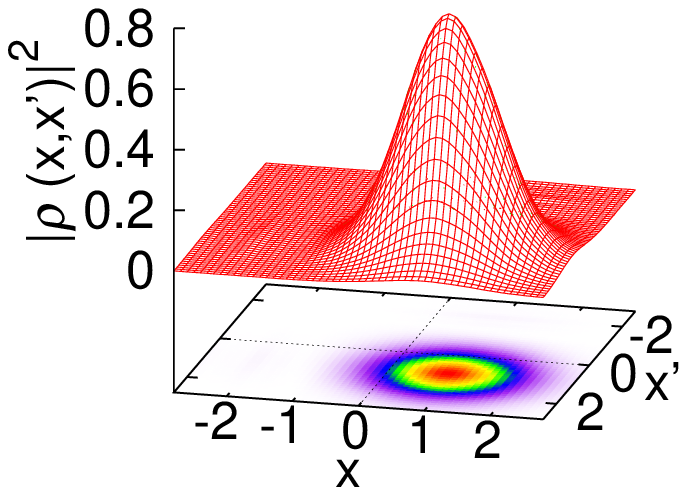}
}
\subfigure[After QND interaction.]{
\includegraphics[clip, scale=0.55]{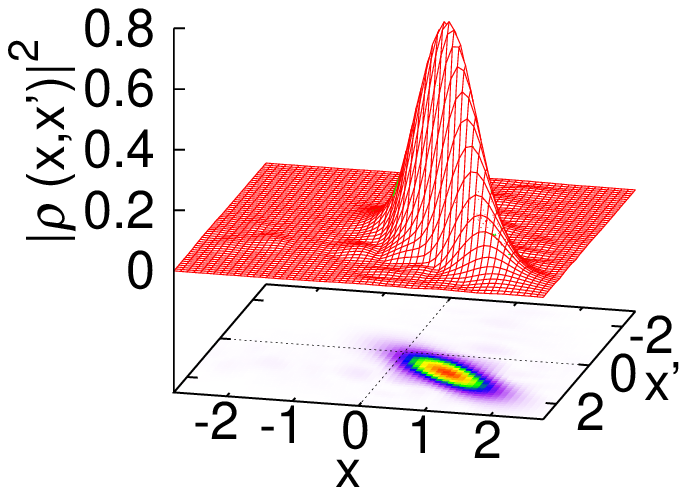}
}
\subfigure[Resulting density matrix after erasing.]{
\includegraphics[clip, scale=0.55]{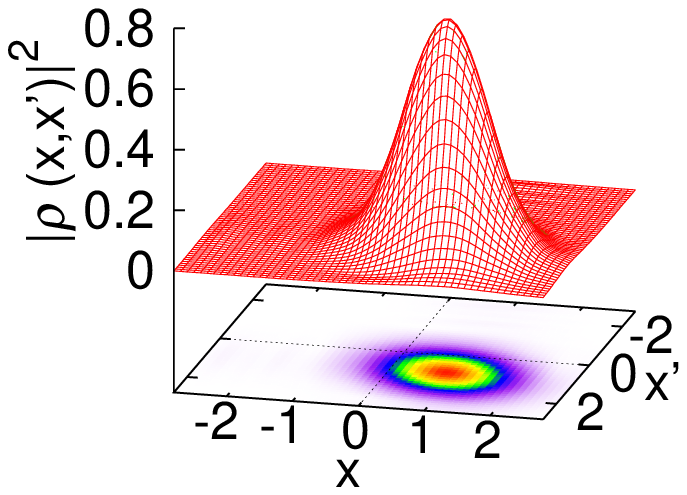}
}
\caption{(Color online) Absolute values of elements of density matrices in each step of the first mode.
These density matrices are reconstructed from experimental marginal distributions shown in Fig.~\ref{fig:erasing_2pi}.
Density matrix of the input coherent state has circular distribution~(a), which shows perfect coherence between different $x$-basis states. The QND interaction with a squeezed vacuum state change it elliptical distribution~(b). Since QND is non-demolition interaction, diagonal ($x=x^\prime$) elements are preserved. Nonetheless, off-diagonal elements are suppressed which shows decrease in coherence. After that coherence is recovered by erasing the leaking information~(c).}
\label{fig:erasing_DM}
\end{figure}

\renewcommand*{\thesubfigure}{(\roman{subfigure})}
\begin{figure}[tb]
\centering
\subfigure[Diagonal elements.]{
\includegraphics[clip, scale=0.4]{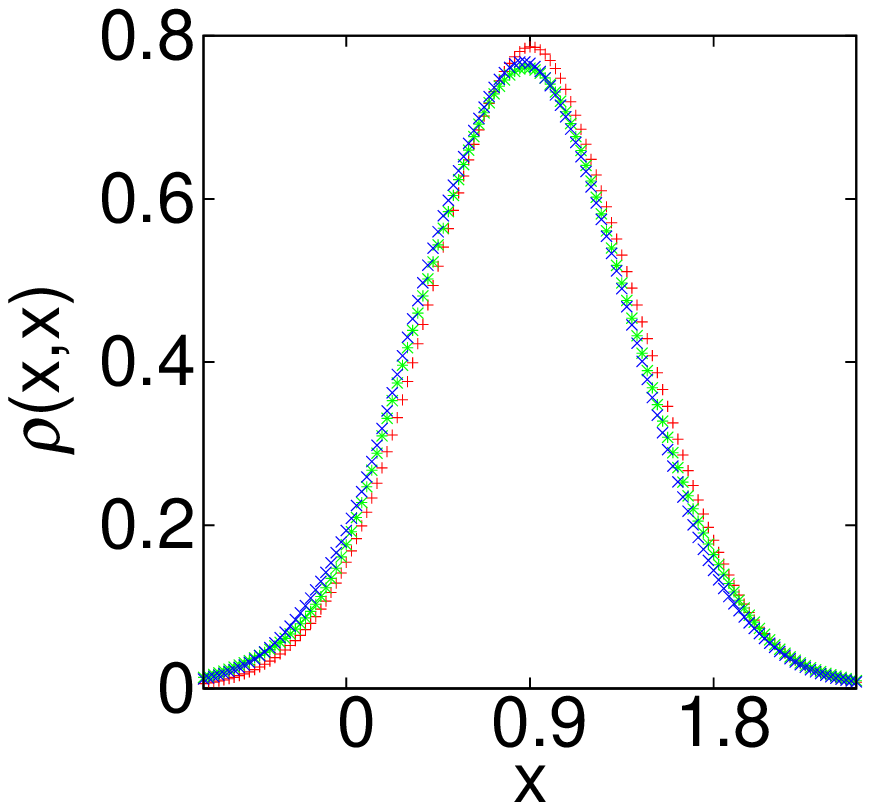}
}
\subfigure[Off-diagonal elements.]{
\includegraphics[clip, scale=0.4]{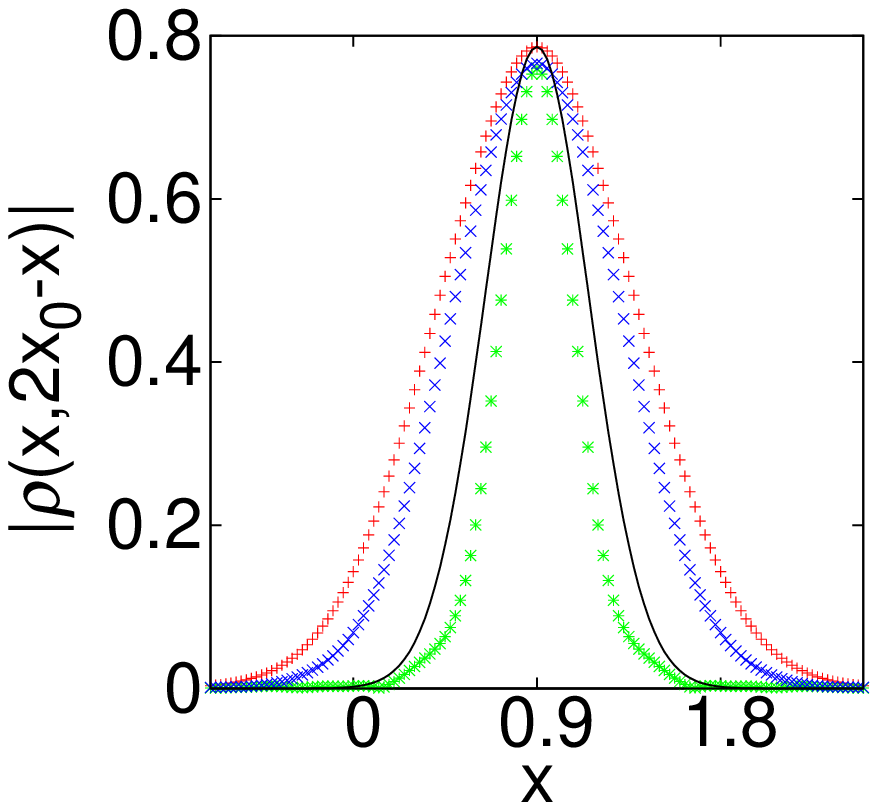}
}
\caption{(Color) Diagonal elements and off-diagonal elements of the density matrix.
Red: input coherent state, Green: after QND interaction, Blue: resulting state after erasing, Black curve: theoretical curve representing maximum coherence after QND interaction with the Fisher information of $I=10.4$.}
\label{fig:erasing_DM_diag}
\end{figure}

\section{Experimental results}
First, we examined quantum states throughout the process examined by homodyne tomography.
Here we utilize a coherent and squeezed-vacuum state as the initial signal and probe state, respectively.
Figure~\ref{fig:erasing_2pi} shows raw data of marginal distributions, while Fig.~\ref{fig:erasing_Wig} and Fig.~\ref{fig:erasing_DM} show reconstructed Wigner functions and density matrices, respectively.
The Wigner functions and the density matrices are reconstructed from the marginal distributions using maximum likelihood method~\cite{Lvovsky09.rmp, Lvovsky04}.
In each of these figures, (a) shows input coherent state and squeezed vacuum state, (b) shows the outcome of the QND interaction, and (c) shows the restored signal coherent state by quantum erasing.

In marginal distributions [Fig.~\ref{fig:erasing_2pi}(a)], input coherent state has a shape of sinusoidal wave with uniform variance corresponding to SNL, while probe squeezed vacuum state has uniformly zero mean amplitude with smaller variance than SNL at the LO phase of $\theta = 0, \pi$ (corresponds to $x$-quadrature) and with larger variance at the phase of $\theta = \pi/2, 3\pi/2$ (corresponds to $p$-quadrature).
These quantum states are shown as circular and elliptical phase space distribution as the Wigner functions shown in Fig.~\ref{fig:erasing_Wig}(a).
After the QND interaction~(b), the mean amplitude and variance of signal $x$-quadrature ($\theta = 0, \pi$) are almost preserved and they are reflected in the probe $x$-quadrature.
These facts clearly show the QND is non-demolishing interaction which transfers $x$-quadrature information from the signal to the probe.
At the same time, the mean amplitude of signal $p$-quadrature ($\theta = \pi/2, 3\pi/2$) is also well-preserved, while the variance is enlarged due to the back action of the information transfer.
This additional variance is originated from the probe $p$-quadrature variance which is preserved during the interaction.
After that, erasing operation is done by measuring probe $p$-quadrature which does not reflect signal information, and by feedforward to the signal.
As the signal resulting state~(c), input coherent state is restored.
Although there is excess noise in $p$-quadrature coming from imperfections of QND interaction, the noise is well suppressed.
The fidelity (overlap between input and output state) is once $F^\stp{b} = \bradmket{\psi^{\stp{a}}}{\rho ^{\stp{b}}}{\psi^{\stp{a}}}=0.47 \pm 0.02$,
which becomes much higher with erasing, $F^\stp{c}=0.86 \pm 0.02$.

We also consider the change in coherence throughout the process.
Figure~\ref{fig:erasing_DM} shows the absolute values of density matrices elements.
Here the diagonal part of each density matrix corresponds to the distribution of $x$, while off-diagonal part corresponds to the coherence between different $x$-basis states.
Input coherent state is $\psi (x) \propto e^{-(x-x_0)^2}e^{i2p_0x}$ as represented in $x$-basis wave function, where $x_0 +ip_0$ is complex amplitude of the coherent state.
In this experiment, the amplitude corresponds to about $0.92+0.90i$.
Density matrix of the input is $\rho(x,x^\prime ) =\psi ^\ast (x^\prime ) \psi (x) \propto e^{-(x-x_0)^2-(x^\prime -x_0)^2}e^{i2p_0(x-x^\prime)}$.
Circular distribution in Fig.~\ref{fig:erasing_DM}(a) agrees well with this equation.
After the QND interaction, diagonal part of the density matrix does not change as shown in Fig.~\ref{fig:erasing_DM}(b), while off-diagonal part is suppressed.
The former corresponds to non-demolitionality of the QND interaction, and the latter corresponds to decoherence.
After erasing, coherence is restored, as shown in Fig.~\ref{fig:erasing_DM}(c).
These changes are much more emphasized in Fig.~\ref{fig:erasing_DM_diag}, where the diagonal elements $\rho (x,x)$ and the off-diagonal elements $\rho (x, 2x_0 -x)$ are cut out from the density matrices.
Note that the change in mean amplitude in Fig.~\ref{fig:erasing_DM_diag}(i) is due to the phase fluctuation during LO phase scan without active feedback.

Next, we perform the erasing operation with exchanging the roles of the first and second modes in order to verify that the back action originating from the probe is removed.
Namely the signal and probe are initially in a squeezed vacuum and coherent states, respectively.
These results are shown in Figs.~\ref{fig:erasing_2pi} and \ref{fig:erasing_Wig} [(a), (b), and (d)] together with the results in previous paragraphs, because the initial state~(a) and the state after the QND~(b) are the same.
After the QND interaction, the $x$-quadrature of the first mode is measured instead of the second mode, and then feedforward to the second mode.
Here, the decoherence occurs on amplitude reflecting the input coherent state, and it disappears after the erasing operation.
The transfer gains from the probe to the signal are about 0.94 in $x$-quadrature before erasing and within $\pm 0.01$ in both quadratures after erasing.
These transfer gains show that the leaking information in the probe is over 98\% erased by erasing.
These results also shows recovery of non-classical property of squeezing in the signal,
the variance of the squeezed quadrature of each step [(a), (b), and (d)] is $-4.9 \pm 0.2$ dB, $1.5 \pm 0.2$ dB and $-1.0 \pm 0.2$ dB relative to the SNL.

\begin{figure}[tb]
\centering
\includegraphics[clip, scale=0.5]{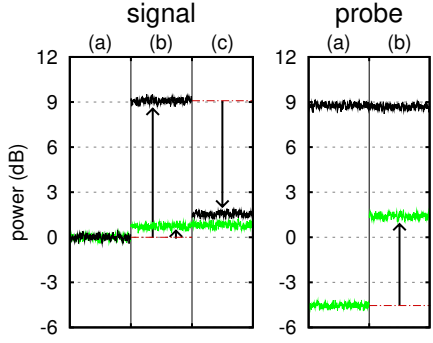}
\caption{(Color online) Experimental results of variances in each step~/~mode of Fig.~\ref{fig:erasing} relative to the shot-noise limit.
Each black trace represents variance of $p$-quadrature while each green trace represents $x$-quadrature.
Initial signal and probe states are a vacuum state and a squeezed vacuum state (a), the signal variance of $x$-quadrature is added to the probe $x$-quadrature while the back action appears in the signal $p$-quadrature owing to a QND interaction (b), the back action is reduced by measuring $p$-quadrature of the probe and performing feedforward (c).}
\label{fig:variances}
\end{figure}

Furthermore, we evaluate variances of input and output states of the QND interaction or the quantum eraser to verify the erasure of {\em leaking information}.
Here, we put a vacuum and squeezed-vacuum states as an initial signal and probe states, respectively, and then we measure the powers with homodyne detectors and a spectrum analyzer.
Figure~\ref{fig:variances} shows experimental results of these variances.
Variance of vacuum state is defined as $1/4 = 0.25$ which corresponds to the SNL.
We obtained the variance of probe-$x$-quadrature $(\Delta x_\RP^\stp{b})^2 = 0.346 \pm 0.006$ after the QND interaction,
the variance of signal-$p$-quadrature $(\Delta p_\RS^\stp{b})^2 = 2.02 \pm 0.03$ after the QND interaction,
and the one $(\Delta p_\RS^\stp{c})^2 = 0.358 \pm 0.005$ after quantum erasing.
Here, $0.25$ of each variance corresponds to the initial vacuum state.
Subtracting this $0.25$ from these variances, we obtain measurement error of a QND measurement $(\Delta \xerr^\stp{b})^2 = 0.096 \pm 0.006$ (corresponding to Fisher information $I=10.4$) and its back action $(\Delta \pba^\stp{b})^2 = 1.77 \pm 0.03$ while quantum erasing suppress the back action quite well to be $(\Delta p_{\mathrm{residual\ noise}}^\stp{c})^2 = 0.108 \pm 0.005$.
Thus, in the case of the QND measurement, inequality \eqref{eq:uncertainty of measurement} is satisfied,
\begin{align}
\Delta x_{\mathrm{error}}^\stp{b} \Delta p_{\mathrm{back\ action}}^\stp{b} &= 0.414 \pm 0.016 \geq \frac{1}{4}.
\end{align}
Note that the uncertainty is not minimum, which is mainly caused by using a mixed state, namely a squeezed thermal state, as the initial probe state.
On the other hand, in the case of quantum erasing, the back action is suppressed below the information erasure criteria of inequality \eqref{eq:uncertainty of measurement2},
\begin{align}
\Delta x_{\mathrm{error}}^\stp{b} \Delta p_{\mathrm{residual\ noise}}^\stp{c} &= 0.102 \pm 0.006 < \frac{1}{4}.
\end{align}
Therefore, the leaking information is successfully erased. The residual Fisher information $I_{\mathrm{residual}}=1.728$ is six-times smaller than before erasing.
Note that the residual noise mainly comes from finitely squeezed ancillas of the QND interaction~\cite{Yoshikawa08.prl,Filip05.pra}.

These relations correspond to the change in coherence shown in Fig.~\ref{fig:erasing_DM_diag}(ii), where the black curve (calculated with Eq.~\eqref{eq:decoherence}) represents maximum coherence with the Fisher information of 10.4.
After the QND interaction, coherence drops below the black curve because of the leaking information.
By erasing, coherence gets restored above the black curve because the leaking information disappears.

\begin{figure}[tb]
\centering
\subfigure[With input amplitude of $1.77$.]{
\includegraphics[clip, scale=0.5]{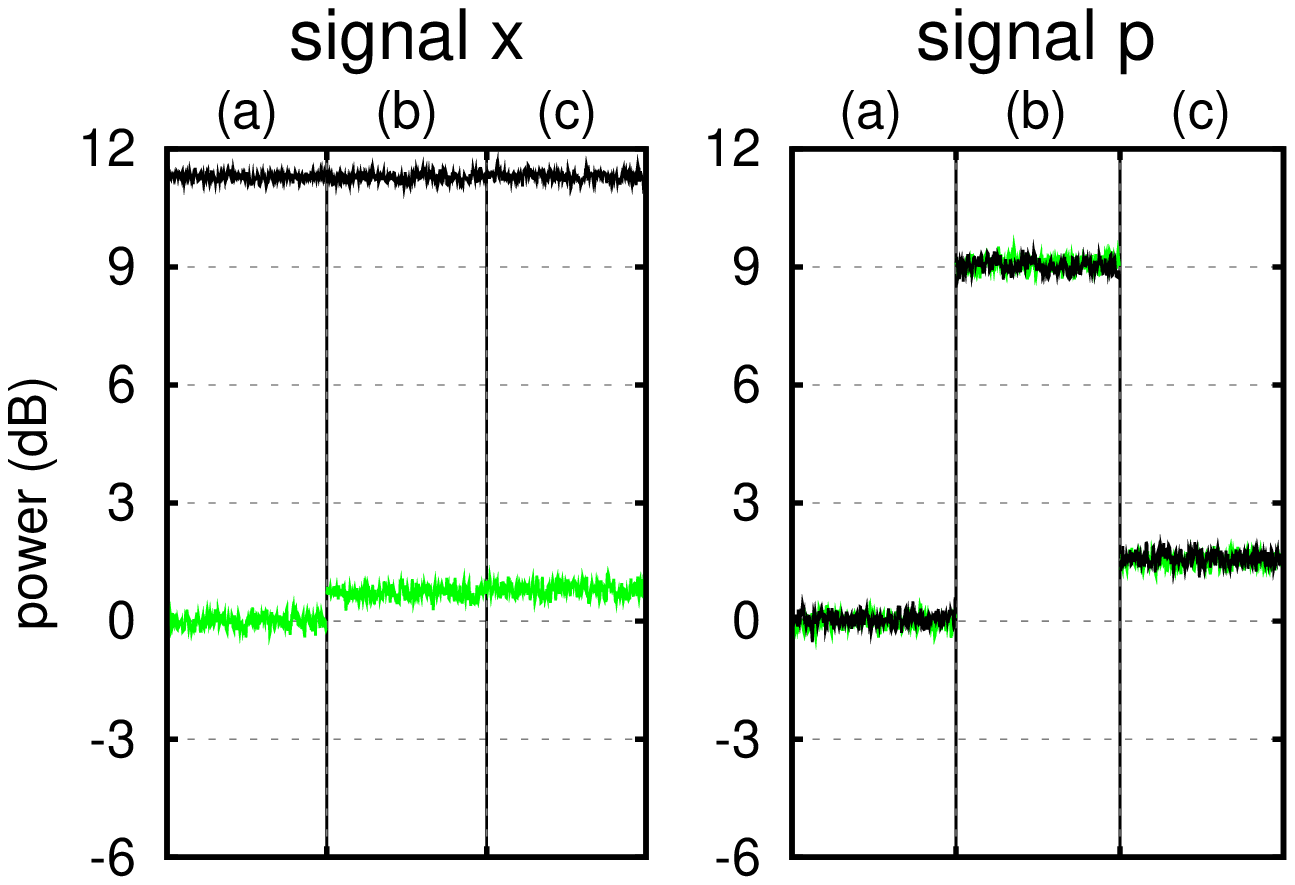}
}
\subfigure[With input amplitude of $i1.39$.]{
\includegraphics[clip, scale=0.5]{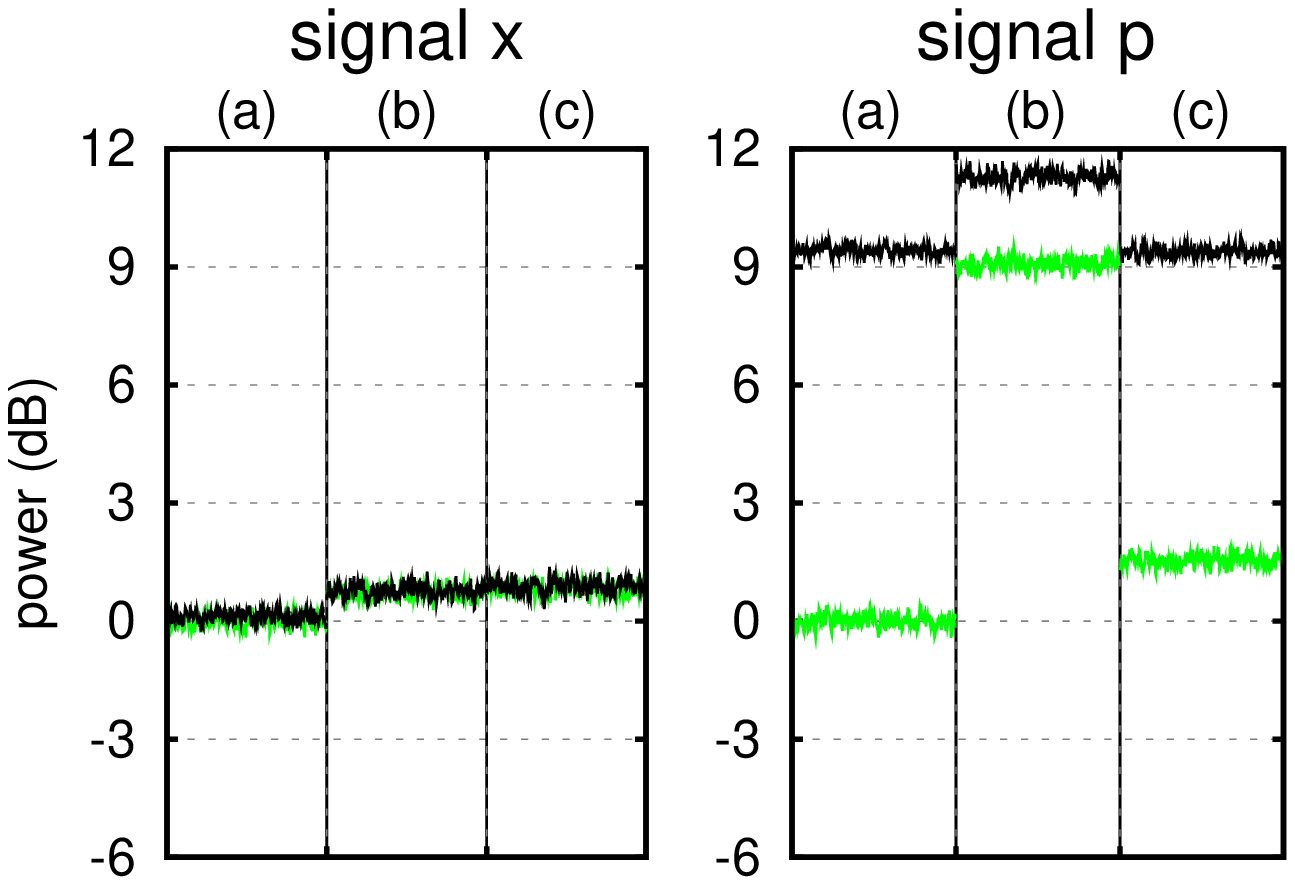}
}
\caption{(Color online) Experimental results of powers in each step~/~quadrature of signal mode of Fig.~\ref{fig:erasing} relative to the shot-noise limit.
Power in each with coherent state input (black trace) is compared with that with vacuum state input (green trace).
The margin of the power corresponds to squared mean amplitude in each quadrature.}
\label{fig:gains}
\end{figure}

Finally, we verify transfer gain and excess noise of our quantum eraser and estimate its average fidelity.
Since both the QND and erasing operation are constructed with Gaussian components (squeezed states as resource, homodyne measurements, and phase space displacements), 
they can be fully characterized by the gains and excess noises in the $x$- and $p$- quadratures, as the components are characterized by those of them.
This Gaussian characterization allows to predict the erasing effect for any input state of the signal.
Coherent states with two different mean complex amplitudes, about $1.77$ and $i1.39$, are utilized as signal input state in order to obtain the gain in $x$- and $p$-quadrature, respectively.
Figure~\ref{fig:gains} shows the input and output powers in both $x$- and $p$- quadratures compared with those of vacuum input.
By the construction of QND interaction, the gains can be adjusted to reach almost unity and the crosstalk between different quadrature can be neglected.
From the results, the gains are calculated as $g_x^\stp{c} = 0.99 \pm 0.01$ and $g_p^\stp{c} = 0.97 \pm 0.01$.
The excess noises are obtained as $(\sigma_x^\stp{c})^2=(\Delta x_\RS^\stp{c})^2-(\Delta x_\RS^\stp{a})^2 = 0.052 \pm 0.004$ and $(\sigma_p^\stp{c})^2=(\Delta p_{\mathrm{residual\ noise}}^\stp{c})^2=0.108 \pm 0.005$.
As the estimated average fidelity, we obtained $F_{\mathrm{avg}}^\stp{c} = 0.862 \pm 0.004$ for a set of coherent states whose mean amplitude is Gaussian-distributed with the variance of $2.5$ i.e. ten times shot noise.
Here we neglect the saturation of detectors or electronic devices because the mean amplitudes are almost the same as aforementioned experiment.
The average fidelity is high compared with that before erasing $F_{\mathrm{avg}}^\stp{b} = 0.442^{+0.009}_{-0.002}$ which calculated with the noise variances of $(\sigma_x^\stp{b})^2 = 0.045 \pm 0.003$ and $(\sigma_p^\stp{b})^2 = 1.77 \pm 0.03$, and with the gains of $g_x^\stp{b} = 0.99 \pm 0.01$ and $g_p^\stp{b} = 0.83^{+0.17}_{-0.01}$
where the range of error on $g_p^\stp{b}$ is enlarged by considering the saturation of detector slew-rate with the sum of antisqueezed variance and mean amplitude.

\renewcommand*{\thesubfigure}{(\alph{subfigure})}
\begin{figure}[tb]
\centering
\subfigure[Verification setup of conditional variance.]{
\includegraphics[clip, scale=0.4]{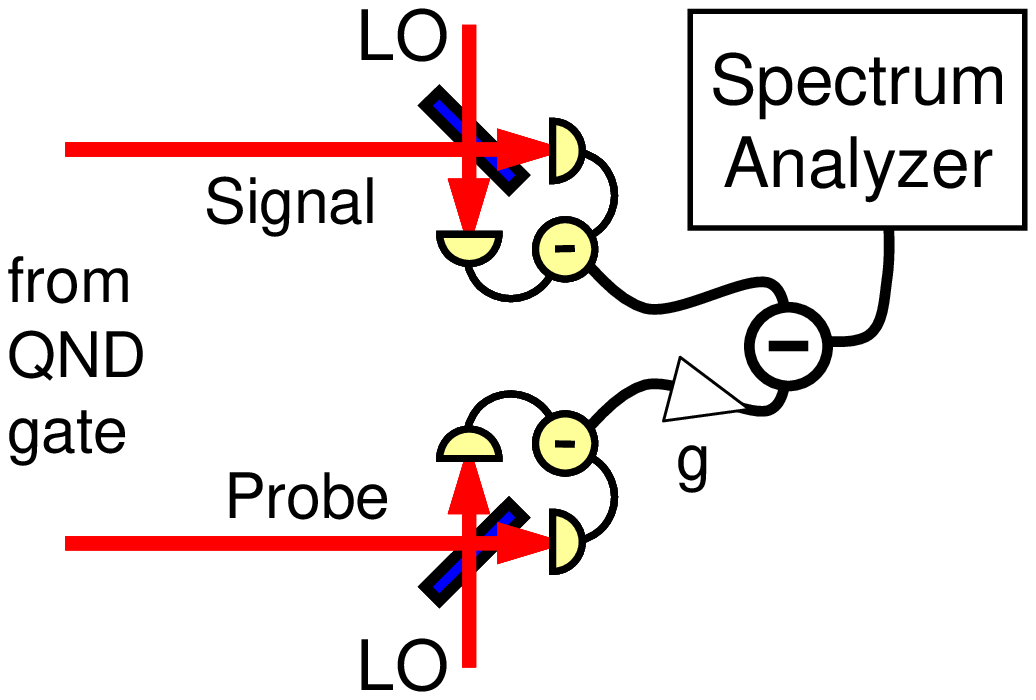}
\label{fig:condv_setup}
}
\subfigure[Conditional variance.]{
\includegraphics[clip, scale=0.4]{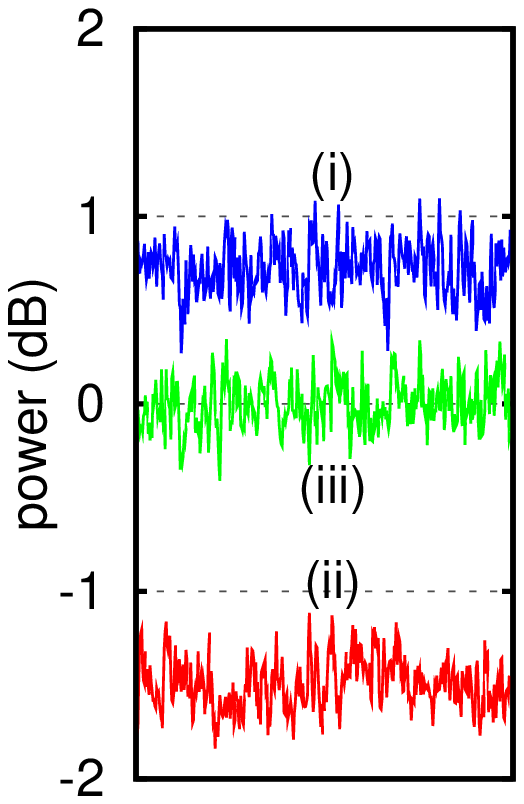}
\label{fig:conditional_variance}
}
\caption{(Color online) Verification setup and experimental results of conditional variance of the signal $x$-quadrature when the probe $x$-quadrature is measured.
Here both the signal and probe are measured via homodyne-detectors, their outcomes are electrically subtracted in the optimal gain $g = 0.56$ and measured by spectrum analyzer.
The signal $x$-quadrature variance [trace (i)] is reduced by measuring $x$-quadrature of probe~(ii), and suppressed below the shot-noise limit~(iii).
}
\end{figure}

For verification of our QND interaction, we also evaluate the performance of QND measurement with the criteria proposed in Ref.~\cite{Holland90}.
Here the input states are the vacuum and squeezed vacuum states again.
As already mentioned, measurement error $(\Delta x_{\mathrm{error}}^\stp{b})^2$ is only $0.096 \pm 0.006$.
The QND variable $x_\RS$ is well preserved within the variance $(\Delta x_\RS^\stp{b})^2 = 0.295 \pm 0.004$ as shown in Fig.~\ref{fig:variances}.
The verification setup represented in Fig.~\ref{fig:condv_setup} yields conditional variance (Fig.~\ref{fig:conditional_variance}),
\begin{align}
\Delta x_{\RS|\RP}^2 &= \min_{g} \expectation{\Delta (\ox_\RS-g\ox_\RP )^2} = 0.177 \pm 0.003 < \frac{1}{4},
\label{eq:conditional variance}
\end{align}
which is suppressed below the SNL.
Thus, there exists nonclassical correlation between the signal and probe quadratures.
These results satisfy the QND-measurement criteria~\cite{Holland90}.
Furthermore, by adjusting subtracting gain $g$ in Eq.~\eqref{eq:conditional variance} to unity i.e.\ $g = 1$, we can also evaluate the entanglement between the signal and probe after the QND interaction.
We obtained the correlation of $x$ quadratures $\Delta (\ox_\RS - \ox_\RP)^2 = 0.243 \pm 0.003 < 1/2$ and the one of $p$ quadratures $\Delta (\op_\RS + \op_\RP)^2 = 0.341 \pm 0.003 < 1/2$,
which satisfy the Duan-Simon entanglement criteria~\cite{Duan00, Simon00}.
Therefore, our QND interaction satisfies the QND and entangling criteria.
So we can conclude that our quantum eraser is undoing such an appropriate QND interaction.


In conclusion, we have experimentally demonstrated CV quantum erasing as reversing a QND interaction.
To verify that the quantum erasing works equally for different input states, we have used a coherent state and a squeezed vacuum state as input states.
We have entangled them by the QND interaction, and then observed that each state is decohered owing to copying the information by the interaction.
These decoherence are observed as decrease of off-diagonal elements of the density matrix.
After that, we have restored either of the two input states.
A full characterization of the quantum erasing as a Gaussian operation have been obtained from estimation of the transfer gains and excess noises.
Initial coherent states has been restored with the fidelity of 86\%, which is verified by two independent ways: homodyne tomography and average fidelity estimation from the transfer gains and excess noises.
A squeezed-vacuum state has also been restored, which shows the erasing operation can recover non-classical properties.
We have verified the erasure of information by using the uncertainty relation between measurement error and back action in conjugate variable, rewritten in terms of the Fisher information and reduction of the off-diagonal elements in the measured basis. By this experimental test, a full analogy between the discrete variable and the continuous variable quantum erasing have been proved.

\section*{Acknowledgement}

This work was partly supported by SCF, GIA, G-COE, PFN and FIRST commissioned by the MEXT of Japan, the Research Foundation for Opt-Science and Technology, SCOPE program of the MIC of Japan, and, 
JSPS and ASCR under the Japan-Czech Republic Research Cooperative Program.
R.~F. acknowledges projects: MSM 6198959213 and ME10156 of the Czech Ministry of Education,
grant 202/08/0224 of GA \v CR and EU Grant FP7 212008 COMPAS.

\appendix
\section{Fisher information and decoherence}
Fisher information is a measure of estimation precision not depending on signal input.
Conditional probability distribution $P(x_0|x_\RS)$ specifies the precision of the estimation of $x_\RS$, where $x_\RS$ and $x_0$ represent input signal coordinate variable and QND measurement outcome, respectively.
By assuming the distribution of measurement error $\xerr(=x_0-x_\RS)$ is independent from $x_\RS$,
The classical Fisher information is defined as follows~\cite{Fisher},
\begin{align}
I&=\int dx_0 \frac{1}{P(x_0|x_\RS )}\left[\frac{\partial P(x_0|x_\RS )}{\partial x_0} \right]^2 .
\end{align}
The Fisher information does not depend on $x_\RS$, we can derive $\rho_\RP(\xerr,\xerr) = P(x_0|x_\RS)$ from Eq.~\eqref{eq:QND2} where $\xerr$ is the measurement error, i.e. $x_0 = x_\RS + \xerr$.
So the Fisher information is the function of the distribution of initial probe coordinate $P(\xerr)=\rho_\RP(\xerr,\xerr)$,
\begin{align}
I=\int d\xerr \frac{1}{P(\xerr )}\left[\frac{\partial P(\xerr )}{\partial \xerr} \right]^2 .
\end{align}
In the case of the initial probe state is Gaussian, i.e.\ $P(\xerr ) = \frac{1}{\sqrt{2\pi}\Delta x^\stp{a}_\RP}\exp \left[ -\frac{\xerr^2}{2(\Delta x^\stp{a}_\RP)^2}\right]$,
the Fisher information is straightforwardly $I=\frac{1}{(\Delta x^\stp{a}_\RP)^2}=\frac{1}{(\Delta \xerr)^2}$.

Then we consider the relation between Fisher information and decoherence.
For example, as we utilized in the experiment, we assume using an $x$-squeezed thermal state as the probe, i.e.\ $\tilde{\rho}_\RP (p,p) \propto \exp \left[ -\frac{p^2}{2(\Delta p^\stp{a}_\RP)^2}\right]$.
The ratio function $R(x^\prime -x)$ is related to the Fisher information, i.e.\ $R(x^\prime -x) = \exp \left[ -2(\Delta p_\RP^\stp{a})^2 (x^\prime -x)^2\right] \leq \exp \left[ -(x^\prime -x)^2/8(\Delta x^\stp{a}_\RP)^2\right]=\exp \left[ -I(x^\prime -x)^2 / 8\right]$,
where we have applied the uncertainty relation $\Delta x^\stp{a}_\RP \Delta p^\stp{a}_\RP \geq 1/4$.
Each element of signal density matrix after the QND interaction satisfies the following,
\begin{align}
|\rho _\RS^\stp{b} (x,x^\prime )|&\leq |\rho_\RS (x,x^\prime )|\exp \left[ -I(x^\prime -x)^2/8 \right] .
\label{eq:decoherence}
\end{align}
As already mentioned, the decoherence exhibits in a reduction of the absolute values $|\rho _\RS^\stp{b} (x,x^\prime )|$ of off-diagonal elements ($x\not= x^\prime$) comparing to original values $|\rho_\RS (x,x^\prime ) |$.
As smaller variance probe we use, the Fisher information $I$ increases, and the decoherence also increases, resulting in a narrowing of the distribution of the off-diagonal elements as shown in Fig.~\ref{fig:erasing_DM_diag}.


\begin{thebibliography}{99}

	\bibitem{Scully.pra}
	M.\ O.\ Scully and Kai Dr\"{u}hl,
	\pra {\bf 25}, 2208 (1982).
%

	\bibitem{trade_off1}
	G.\ Jaeger, A.\ Shimony, and L.\ Vaidman, \pra {\bf 51}, 54 (1995).
	
	\bibitem{trade_off2}
	B.-G.\ Englert, \prl {\bf 77}, 2154 (1996).

	\bibitem{Kim99.prl}
	Y.-H.\ Kim, R.\ Yu, S.\ P.\ Kulik, Y.\ Shin, and M.\ O.\ Scully,
	\prl {\bf 84}, 1 (1999).

    \bibitem{Barbieri.njp}
	M.\ Barbieri, M.\ E.\ Goggin, M.\ P.\ Almeida, B.\ P.\ Lanyon and A.\ G.\ White
    New J. Phys. {\bf 11}, 093012 (2009).
	
	\bibitem{Radim02.ipp}
	R.\ Filip, J.\ Opt.\ B {\bf 4}, 202 (2002).
	
	\bibitem{Filip03.pra}
	R.\ Filip,
	\pra {\bf 67}, 042111 (2003).
	
	\bibitem{Schwindt}
	P.\ D.\ D.\ Schwindt, P.\ G.\ Kwiat, and B.-G.\ Englert, \pra {\bf 60}, 4285 (1999).

	\bibitem{cnot}
	M.A. Nielsen and I.L. Chuang,
    Quantum Computation and Quantum Information, Cambridge University Press, ISBN 0-521-63235-8 (2000).
	
	\bibitem{Filip05.pra}
	R.\ Filip, P.\ Marek, and U.L.\ Andersen,
	\pra {\bf 71}, 042308 (2005).

	\bibitem{Yoshikawa08.prl}
	J.\ Yoshikawa, Y.\ Miwa, A.\ Huck, U.\ L.\ Andersen, P.\ van Loock, and A.\ Furusawa,
	\prl {\bf 101}, 250501 (2008).

    \bibitem{Andersen04.prl}
	U.\ L.\ Andersen, O.\ Gl\"ockl, S.\ Lorenz, G.\ Leuchs, and R.\ Filip,
	\prl {\bf 93}, 100403 (2004).
	
	\bibitem{Filip09}
    R.\ Filip, \pra {\bf 80}, 022304 (2009).
	
	\bibitem{Sabuncu}
	M.\ Sabuncu, R.\ Filip, and G.\ Leuchs, and U.\ L.\ Andersen, \pra {\bf 81}, 012325 (2010).
	
	\bibitem{shaping}
	Y.\ Miwa, R.\ Ukai, J.\ Yoshikawa, R.\ Filip, P.\ van Loock, and A.\ Furusawa,
	\pra {\bf 82}, 032305 (2010).
	
	\bibitem{one-wayGu}
	M.\ Gu, C.\ Weedbrook, N.\ C.\ Menicucci, T.\ C.\ Ralph, and P.\ van Loock,
	\pra {\bf 79}, 062318 (2009).
	
    \bibitem{Fisher}
    R. A. Fisher, Proc. Cambridge Phil. Soc. 22, 700 (1929) reprinted in
    Collected Papers of R. A. Fisher, edited by J. H. Bennett (Univ. of
    Adelaide Press, South Australia, 1972), pp. 15-40], for a recent overview, B. R. Frieden, {\em Science from Fisher Information}, Cambridge University Press, Cambridge, England, 2004.

    \bibitem{Braginsky}
 	V.\ B.\ Braginsky and F.\ Y.\ Khalili, {\it Quantum measurement} (Cambridge University Press, Cambridge, 1992).
	
	
	\bibitem{Miwa09.pra}
	Y.\ Miwa, J.\ Yoshikawa, P.\ van Loock, and A.\ Furusawa,
	\pra {\bf 80}, 050303(R) (2009).

    \bibitem{Suzuki06.apl}
    S.\ Suzuki, H.\ Yonezawa, F.\ Kannari, M.\ Sasaki, and A.\ Furusawa,
    Appl. Phys. Lett. {\bf 89}, 061116 (2006).

	\bibitem{Yoshikawa07.pra}
	J.\ Yoshikawa, T.\ Hayashi, T.\ Akiyama, N.\ Takei, A.\ Huck, U.L.\ Andersen, and A.\ Furusawa,
	\pra {\bf 76}, 060301(R) (2007).
	
	\bibitem{Lvovsky09.rmp}
	A.\ I.\ Lvovsky and M.G. Raymer, \rmp {\bf 81}, 299 (2009).
	
	\bibitem{Lvovsky04}
	A.\ I.\ Lvovsky, J.\ Opt.\ B {\bf 6}, S556 (2004).  
	
	\bibitem{Holland90}
	M.\ J.\ Holland, M.\ J.\ Collett, and D.\ F.\ Walls, and M.\ D.\ Levenson, \pra {\bf 42}, 2995 (1990).
	
	\bibitem{Duan00}
	L.-M.\ Duan, G.\ Giedke, J.I.\ Cirac, and P.\ Zoller,
	\prl {\bf 84}, 2722 (2000).

	\bibitem{Simon00}
	R.\ Simon, \prl {\bf 84}, 2726 (2000).
	
	
\end{thebibliography}
\end{document}